\documentclass[12pt]{article}

\ifx\pdfoutput\undefined
\usepackage[dvips,bookmarks]{hyperref}
\else
\usepackage{hyperref}
\fi
\hypersetup{colorlinks=false,bookmarksopen,bookmarksnumbered,citecolor=blue,
   pdfstartview=FitH}

\usepackage{latexsym}
\usepackage{amssymb,amsfonts,amsmath}
\usepackage{graphicx} 
\usepackage{indentfirst}
\usepackage{tensor}
\usepackage{bbm}
\usepackage{slashed, tensor}

\parskip=\medskipamount

\usepackage[mathscr]{euscript} 

\newcommand {\cC}{{\cal C}}

\newcommand {\cE}{{\cal E}}
\newcommand {\cF}{{\cal F}}

\newcommand {\cL}{{\cal L}}
\newcommand {\cM}{{\cal M}}
\newcommand {\cN}{{\cal N}}

\newcommand {\cR}{{\cal R}}

\newcommand {\cT}{{\cal T}}

\newcommand {\cV}{{\cal V}}
\newcommand {\cW}{{\cal W}}

\newcommand {\cZ}{{\cal Z}}

\def\a{\alpha}
\def\b{\beta}
\def\c{\chi}
\def\d{\delta}

\def\g{\gamma}
\def\G{\Gamma}

\def\l{\lambda}

\def\q{\theta}

\def\s{\sigma}

\def\D{\Delta}

\def\L{\Lambda}

\def\S{\Sigma}

\def\X{\Xi}

\def\ri{{\rm i}}
\def\re{{\rm e}}

\newcommand{\gd}{{\dot\g}}
\newcommand{\dd}{{\dot\d}}
\newcommand{\ad}{{\dot{\alpha}}}
\newcommand{\bd}{{\dot{\beta}}}
\newcommand{\dalpha}{{\dot{\alpha}}}
\newcommand{\dbeta}{{\dot{\beta}}}

\newcommand{\1}{{\underline{1}}}
\newcommand{\2}{{\underline{2}}}

\newcommand{\ve}{\varepsilon}

\newcommand{\hf}{\frac12}

\newcommand{\psib}{\bar{\psi}}

\newcommand{\be}{\begin{equation}}
\newcommand{\ee}{\end{equation}}
\newcommand{\bea}{\begin{eqnarray}}
\newcommand{\eea}{\end{eqnarray}}
\newcommand{\non}{\nonumber}
\newcommand{\ba}{\begin{array}}
\newcommand{\ea}{\end{array}}

\newcommand{\bm}[1]{\mbox{\boldmath$#1$}}

\def\double #1{#1{\hbox{\kern-2pt $#1$}}}

\newcommand{\ts}{{\tilde{\s}}}

\newcommand{\bsubeq}{\begin{subequations}}
\newcommand{\esubeq}{\end{subequations}}

\newcommand{\rd}{\mathrm d}
\newcommand{\HC}{{\mathrm{c.c.}}}

\newcommand{\bphi}{{\bar\phi}}
\newcommand{\bpsi}{{\bar\psi}}

\newcommand{\eps}{{\epsilon}}

\newcommand{\eol}{\notag \\}
\newcommand{\lc}{{\vert}}

\newcommand{\poin}[1]{\hat{#1}}

\numberwithin{equation}{section}  

\renewcommand{\eps}{\ve}

\begin{document}
\begin{titlepage}
\begin{flushright}
October 2012\\
\end{flushright}

\begin{center}
{\Large \bf 
Variant vector-tensor multiplets in supergravity: Classification and component reduction
}
\end{center}

\begin{center}

{\bf
Joseph Novak} \\
\vspace{5mm}

\footnotesize{
{\it School of Physics M013, The University of Western Australia\\
35 Stirling Highway, Crawley W.A. 6009, Australia} \\
\vspace{2mm}
{\it Email:} \texttt{joseph.novak@uwa.edu.au}
}  
~\\
\vspace{2mm}

\end{center}

\begin{abstract}
\baselineskip=14pt

The recent paper arXiv:1205.6881 has developed superform formulations for two versions of the vector-tensor multiplet and their 
Chern-Simons couplings in four-dimensional $\cN = 2$ conformal supergravity. One of them is the standard vector-tensor multiplet 
with the central charge gauged by a vector multiplet. The other is the variant vector-tensor multiplet with the property that its own one-form gauges the central charge. 
Here a more general setup is presented in which the known versions reside as special cases. Analysis of the setup demonstrates that under certain assumptions 
there are two distinct variants, corresponding to the two formulations in arXiv:1205.6881. This provides a classification 
scheme for vector-tensor multiplets.

We then show that our superspace description leads to an efficient means of deriving component actions in supergravity. The entire action including all fermionic terms is 
derived for the non-linear vector-tensor multiplet. This extends the results of de Wit {\it et al.} in hep-th/9710212, where only the bosonic sector appeared. Finally, the
bosonic sector of the action for the variant vector-tensor multiplet is given.

\end{abstract}

\vfill
\end{titlepage}

\newpage
\renewcommand{\thefootnote}{\arabic{footnote}}
\setcounter{footnote}{0}

\tableofcontents{}
\vspace{1cm}
\bigskip\hrule


\section{Introduction}

In $\cN = 2$ supersymmetry there exist off-shell supermultiplets requiring the presence of a central charge, such as the Fayet-Sohnius hypermultiplet \cite{Fayet, Sohnius}. 
The vector-tensor (VT) multiplet \cite{SSW, SSW2} is another such supermultiplet that has received a great deal of interest due to its significance in the context of string compactifications \cite{deWKLL}.\footnote{From the 
point of view of $\cN = 1$ supersymmetry the multiplet decomposes into a vector and a tensor multiplet \cite{Milewski}.} 
Its physical fields consist of a real scalar, a doublet of Weyl spinors and a gauge one-form and two-form. The multiplet may be viewed as a
dual realization of the $\cN = 2$ Abelian vector multiplet, obtained by dualizing one of the two physical scalars of the vector multiplet into a gauge two-form.

Numerous papers have been devoted to the study of the VT multiplet and its Chern-Simons couplings, both in the component approach with the use of superconformal tensor 
calculus \cite{Claus1, Claus2, Claus3} and in the framework of conventional $\cN = 2$ superspace \cite{HOW, GHH, BHO} as well as $\cN = 2$ harmonic superspace 
\cite{DKT, DK, IS, DIKST}. In particular, de Wit {\it et al.} \cite{Claus3} provided a comprehensive analysis of the VT multiplet and its Chern-Simons couplings in supergravity, although 
due to its complexity only the bosonic sector of the action was computed. This demonstrated the need for a suitable superspace formulation. An important step 
towards such a formulation was made by \cite{DIKST}, 
where a general setting for $\cN =2$ rigid supersymmetric theories with gauged central charge was developed using harmonic superspace. Only recently a superfield formulation for the VT multiplet in 
supergravity has been found \cite{KN, BN, Novak}.\footnote{Here we are concerned with off-shell multiplets, however see \cite{ADS, ADST} for a different approach using Free Differential Algebra.} This delay was 
due in part to the absence of sufficiently simple superspace descriptions for supergravity, which has appeared in recent years \cite{KLRT-M08, Butter4D}.

In addition to the standard VT multiplet, it was discovered by Theis \cite{Theis1, Theis2} that in rigid supersymmetry there exists a {\it variant} VT multiplet.\footnote{In \cite{Theis1, Theis2} the 
variant VT multiplet was referred to as the new non-linear VT multiplet.} This multiplet is in stark contrast to both the linear \cite{SSW, SSW2} and non-linear \cite{Claus1} forms of the standard 
VT multiplet. It is associated with a new procedure of gauging the central charge, with the multiplet itself providing the central charge gauge one-form. This 
is unlike the standard gauging which makes use of a vector multiplet. The variant VT multiplet has since been generalized 
to supergravity \cite{Novak}.

In rigid supersymmetry, the difference between the versions of the VT multiplet may be illustrated in $\cN = 2$ central charge superspace \cite{Sohnius} in 
which the spinor covariant derivatives obey the anti-commutation relations
\begin{subequations}
\begin{align}
\{ D_\a^i, D_\b^j \} &= - 2 \eps_{\a\b} \eps^{ij} \D  ~, \quad \{ \bar{D}^\ad_i, \bar{D}^\bd_j \} = 2 \eps^{\ad\bd} \eps_{ij} \D ~, \\
\{ D_\a^i , \bar{D}^\bd_j \} &= -2 \ri \d^i_j \partial_\a{}^\bd ~,
\end{align}
\end{subequations}
where $\D$ is the central charge. The VT multiplets correspond to different constraints on a real superfield, $L$. For instance, the original {\it linear} VT multiplet \cite{SSW, SSW2} can be 
described by the constraints \cite{GHH, DKT}
\begin{align} D^{ij} L = 0 \ , \quad 
D_\a^{(i} \bar{D}_\ad^{j)} L = 0 \ ,
\end{align}
where $D^{ij}  :=  D^{\a (i} D_{\a}^{j)}$ and $\bar D_{ij} := \bar D_{\ad (i} \bar D^\ad_{j)}$.\footnote{The multiplet is on-shell if $\D L = 0$.} It is worth mentioning that these constraints may be formulated 
within the superform formulations of \cite{HOW, GHH, BHO}.

Variant versions of the linear VT multiplet can be constructed by looking for consistent deformations of its constraints \cite{DK, IS, DIKST}. There are 
two possible distinct deformations (up to superfield redefinitions):
\begin{subequations}
\begin{align}
{\rm (i)} \quad D^{ij} L &= - \frac{1}{L} D^i L D^j L + \frac{1}{L} \bar{D}^i L \bar{D}^j L \ , \quad D_\a^{(i} \bar{D}_\ad^{j)} L = 0 \ , \\
{\rm (ii)} \quad D^{ij} L &= 2 \tan(2 L) D^i L D^j L + \frac{2}{\cos(2 L)} \bar{D}^i L \bar{D}^j L \ , \quad D_\a^{(i} \bar{D}_\ad^{j)} L = 0 \ .
\end{align}
\end{subequations}
The first deformation was given in \cite{DK, IS} and at the component level corresponds to the {\it non-linear} VT multiplet of \cite{Claus1}. The self-interaction for (i) may be 
attributed to the presence of a Chern-Simons form.  Deformation (ii) was found in \cite{IS}, where it was shown that its three-form field strength cannot be locally solved in terms of a 
two-form gauge potential. This was why (ii) was not considered in the earlier work \cite{DK}.\footnote{S. Kuzenko, {\it private communication}.} However, later it was demonstrated 
by Theis \cite{Theis1, Theis2} that the one-form gauge potential corresponding to (ii) can be used to gauge the central charge, and in this way an appropriate two-form gauge potential may be defined.

In light of the three different versions of the VT multiplet in rigid supersymmetry, it is natural to ask whether they may be lifted to supergravity. Remarkably it turns out that the linear VT multiplet 
does not generalize to conformal supergravity nor to $\rm AdS$ \cite{KN}, with supergravity generalizations only existing for the deformed versions.\footnote{Although 
there exists a linear case of the VT multiplet in \cite{Claus3, KN} requiring an additional vector multiplet. It should be clear from context to which multiplet 
we are referring to.} In supergravity the non-linear VT multiplet together with its general couplings was originally addressed in \cite{Claus3} at the component 
level and at the superfield level in \cite{KN, BN}, with \cite{Novak} elucidating the origin of the constraints.\footnote{A linear case can exist in supergravity provided it is coupled to 
two vector multiplets \cite{Claus3, KN}. However it may be viewed as a special case of the non-linear VT multiplet with Chern-Simons terms after a superfield redefinition \cite{BN}.} 
Moreover in \cite{Novak} the variant VT multiplet together with its couplings 
to vector multiplets was constructed by using a completely superform formulation.

At present there exist only two versions of the VT multiplet within $\cN = 2$ conformal supergravity; the non-linear VT multiplet and the variant VT multiplet. However, there still remains the possibility 
of alternative off-shell realizations in supergravity. For example, in \cite{BKN} it was shown that it is possible to use the new procedure for 
gauging the central charge to construct new off-shell formulations for the {\it massless} Fayet-Sohnius hypermultiplet. Furthermore it would be interesting if supergravity offers new versions 
not available in the rigid supersymmetric case. This paper endeavors to examine this issue.

After having described VT multiplets through their superfield constraints (ensuring the appropriate component structure), an action is necessary. The linear 
multiplet \cite{BS} has become a powerful tool for the construction of action functionals for supergravity-matter systems, especially after its reformulation within superconformal tensor calculus \cite{deWvHVP3}. 
In particular, in \cite{Claus3} the linear multiplet was used to derive the bosonic sector of the action for the standard VT multiplet.

A locally supersymmetric action based on the linear multiplet was constructed in harmonic superspace in \cite{KT}. The harmonic superspace approach to $\cN = 2$ supergravity was developed in terms of certain 
prepotentials \cite{HarmSpace}. However it is 
not known how to relate the prepotentials to superspace differential geometry given in \cite{Howe, KLRT-M08, Butter4D}. This renders it impractical for our consideration, since both the constraints and the superspace 
Lagrangian \cite{KN, BN} for the standard VT multiplet involve covariant derivatives. Recently, the suitable superspace action was constructed in \cite{BKN}. It makes use of the ectoplasm approach \cite{Ectoplasm, GGKS}, 
also known as the superform approach for constructing supersymmetric invariants.\footnote{The mathematical formalism behind the 
ectoplasm approach \cite{Ectoplasm, GGKS} is a special case of the theory of integration over surfaces in supermanifolds, see for instance \cite{Vor}. The idea behind the ectoplasm approach 
is to use a closed superform. In four-dimensions, to the best of our knowledge this idea was first suggested by Hasler \cite{Hasler}.}  
One of the main results of \cite{BKN} was a new action principle based on 
a {\it deformed} linear multiplet, corresponding to the case where the central charge one-form potential is not annihilated by the central charge. This is precisely the case 
for the variant VT multiplet whose superfield Lagrangian was given in \cite{Novak}. It would therefore be interesting to derive new 
component results entirely from superspace. This is the other main goal of this paper.

This paper is structured as follows. In section \ref{GCC} we describe a general procedure to gauge a real central charge in $\cN = 2$ conformal supergravity. Section \ref{superform} is devoted to 
the general setup and superform formulation for VT multiplets. The general analysis will lead us to a set of constraints and possibilities will be analyzed. In section \ref{compActions}, building on 
the methods of \cite{BN}, we provide component actions for two types of VT multiplets. Section \ref{discus} concludes the paper. Two technical appendices are also included. 
Appendix \ref{conformalSpace} contains a summary of the conformal supergravity formulation of \cite{Butter4D} used in this paper and appendix \ref{CompIdentities} contains useful component 
results derived from the superspace formulation of \cite{Butter4D}.


\section{Gauging the central charge in $\cN = 2$ conformal supergravity} \label{GCC}

Off-shell formulations for VT multiplets in supergravity require the presence of a one-form to gauge the central charge. It is well known that for this purpose one can use the one-form of an 
off-shell vector multiplet. However, this implies that the gauge connection is annihilated by the central charge. There is a more general approach \cite{Novak}, inspired by 
the construction of Theis \cite{Theis1, Theis2}, which may be used to describe the variant VT multiplet. In this section we review the approach of \cite{Novak}. We make use of the 
superspace formulation for $\cN = 2$ conformal supergravity developed in \cite{Butter4D} as reformulated in \cite{BN} (see appendix \ref{conformalSpace}).\footnote{All results derived within the 
formulation of \cite{Butter4D} may be extended via gauge fixing to the formulations of \cite{Howe} and \cite{KLRT-M08}.}


\subsection{Setup}

Starting with $\cN = 2$ conformal superspace \cite{Butter4D}, we introduce a real central charge, $\D$, which we require to obey the Leibniz rule and commute with all superconformal 
generators and covariant derivatives\footnote{The central charge may be thought of as a derivative with respect to an extra bosonic coordinate as in \cite{BKN}.}
\be [\D, \nabla_A] = 0 \ .
\ee
To gauge the central charge we define the gauge covariant derivatives
\be \bm \nabla_A := \nabla_A + \cV_A \D \ , \quad \bm \nabla := E^A \bm \nabla_A \ ,
\ee
where gauge connection $\cV := E^A \cV_A$ is not assumed to be annihilated by the central charge ({\it i.e.} we may have $\D \cV_A \neq 0$). Then the algebra of 
gauge covariant derivatives is
\begin{align} \label{GCDA_sep}
[\bm \nabla_A, \bm \nabla_B\} &= [\nabla_A, \nabla_B\} + \cF_{AB} \D \ , \non\\
[\D , \bm \nabla_A] &= \cF_{zA} \D := \D \cV_A \D \ ,
\end{align}
where $\cF = \hf E^B E^A \cF_{AB}$ is the field strength
\be
\cF = \bm \nabla \cV \ , \quad \cF_{AB} = 2 \bm \nabla_{[A} \cV_{B \} } - T_{AB}{}^C \cV_C \ .
\ee

It proves advantageous to treat $\D$ on the same footing as the covariant derivatives $\bm \nabla_A$.\footnote{See \cite{AGHH} for a similar treatment in supergravity.} To do this we introduce the notation
\be \hat{\bm \nabla}_{\hat{A}} := (\bm \nabla_A, \bm \nabla_z := \D)
\ee
and write the algebra of gauge covariant derivatives \eqref{GCDA_sep} as
\begin{align}
[\hat{\bm \nabla}_{\hat{A}}, \hat{\bm \nabla}_{\hat{B}} \} &= T_{\hat{A}\hat{B}}{}^{\hat{C}} \bm \nabla_{\hat{C}} + \hf R_{\hat{A} \hat{B}}{}^{cd} M_{cd} + R_{\hat{A} \hat{B}}{}^{kl} J_{kl}
	\eol & \quad
	+ \ri R_{\hat{A}\hat{B}}(Y) Y + R_{\hat{A}\hat{B}} (\mathbb{D}) \mathbb{D} + R_{\hat{A}\hat{B}}{}^C K_C \ ,
\end{align}
where we define
\begin{align}
T_{AB}{}^z := \cF_{AB} \ , \quad T_{A z}{}^B = 0
\end{align}
and all curvatures involving a $z$ index vanish. Finally the Bianchi identities are
\be \hat{\bm \nabla}_{[\hat{A}} \hat{\bm \nabla}_{\hat{B}} \hat{\bm \nabla}_{\hat{C}\}} = 0 \quad \rightarrow \quad \hat{\bm \nabla}_{[\hat{A}} \cF_{\hat{A} \hat{B}\}} - T_{[\hat{A} \hat{B}}{}^{\hat{D}} \cF_{|\hat{D}| \hat{C} \} } = 0 \ ,
\ee
where $\cF_{\hat{A} \hat{B}} := (\cF_{AB}, \cF_{zA})$ may be interpreted as a field strength.

Local central charge 
transformations are realized on the gauge covariant derivatives and tensor superfields, $U$, as
\begin{align}
\d_\L \bm \nabla_A &= [\L \D, \bm \nabla_A] \quad \Leftrightarrow \quad \d_\L \cV_A = \L \D \cV_A - \nabla_A \L \ , \non\\
\d_\L U &= \L \D U \ , 
\end{align}
where the vielbein and superconformal connections are all inert under central charge transformations and $\D \L = 0$.\footnote{The central charge must annihilate 
$\L$ to ensure the central charge descendant $\D U$ transforms covariantly under central charge transformations.} It follows that the field strength transforms covariantly
\be \d_\L \cF_{\hat{A}\hat{B}} = \L \D \cF_{\hat{A}\hat{B}} \ .
\ee

We can see that the Bianchi identities on the original superspace curvatures are unaffected by the gauging procedure and they are constrained in the usual way. However, in order 
to describe a background multiplet with $8+8$ off-shell degrees of freedom we must constrain the field strength $\cF_{\hat{A} \hat{B}}$ in some way, which we now turn to.


\subsection{The large vector multiplet}

Following \cite{Novak} we impose constraints at mass dimension-1 on the field strength that resemble those for a standard vector multiplet \cite{GSW}
\be \cF_\a^i{}_\b^j = - 2 \eps_{\a\b} \eps^{ij} \bar{M} \ , \quad \cF^\ad_i{}^\bd_j = 2 \eps^{\ad\bd} \eps_{ij} M \ , \quad \cF_\a^i{}^\bd_j = 0 \ ,
\ee
except that $M$ is not required to be annihilated by the central charge.\footnote{The 
definition of $M$ differs by a factor of $\ri$ from \cite{Novak} and by a minus sign from \cite{BKN}.} The superfield $M$ must be conformally primary with dimension 1 and $\rm U(1)$ 
charge -2 (a Jacobi identity implies that $M$ is conformally primary \cite{Butter4D})
\be K_A M = 0 \ , \quad \mathbb D M = M \ , \quad Y M = -2 M \ .
\ee
The Bianchi identities on the field strength then lead to the constraints
\begin{gather} \label{eq_MBianchi1}
\bm\nabla_\alpha^{(i} \bar{\bm\nabla}_\dalpha^{j)} \ln \left(\frac{M}{\bar M}\right) = 0~, \\
\bar M \bm\nabla^{ij} \left(\frac{M}{\bar M} \right) = 
M \bar {\bm\nabla}^{ij} \left(\frac{\bar M}{M} \right) \label{eq_MBianchi2} \ ,
\end{gather}
where
\be \bm \nabla^{ij} := \bm \nabla^{\a (i} \bm \nabla^{j)}_\a \ , \quad \bar{\bm \nabla}^{ij} := \bar{\bm \nabla}^{(i}_\ad \bar{\bm \nabla}^{j) \ad} \ .
\ee
The remaining components of $\cF$ are
\begin{subequations}
\begin{align}
\cF_z{}_\alpha^i &= \bm\nabla_\alpha^i \ln \bar M~, \\
\cF_a{}_\beta^j &= \frac{\ri}{2} (\sigma_a)_\beta{}^\dalpha M
	\bar{\bm\nabla}_\dalpha^j \left(\frac{\bar M}{M}\right)~, \\
\cF_{za} &= -\frac{\ri}{8} (\sigma_a)_{\alpha \dalpha}
	(\bm\nabla^{\alpha k} \bar{\bm\nabla}_k^\dalpha \ln M
	+ \bar{\bm\nabla}_k^\dalpha \bm\nabla^{\alpha k} \ln \bar M)~, \\
\cF_{ab} &= - \frac{1}{8} (\sigma_{ab})^{\alpha \beta} 
	(\bar M \bm\nabla_{\alpha \beta} \Big(\frac{M}{\bar M}\Big) + 4 \bar M W_{\alpha \beta}) + \HC
\end{align}
\end{subequations}
It is interesting to note that the expression for $\cF_{z}{}_\a^i$ implies that the operator $\bar{M} \D$ commutes with $\bm \nabla_\a^i$
\begin{align}
[\bar M \Delta, \bm\nabla_\alpha^i] = 0 ~, \qquad
[M \Delta, \bar{\bm\nabla}^\dalpha_i] = 0~.
\end{align}

The resulting multiplet is called the large vector multiplet, due to the fact that it contains too many component fields to correspond to a multiplet of $8+8$ degrees of freedom. 
Therefore it is necessary to constrain the component fields of $M$ in addition to the constraints \eqref{eq_MBianchi1}. There are two known choices of significance.

One obvious choice is to let $M$ be independent of the central 
charge, which leads us to $M = \cZ$ with $\cZ$ a vector multiplet
\be
\bar{\nabla}_{\ad}^i \cZ = 0~, \quad \nabla^{ij} \cZ = \bar{\nabla}^{ij} \bar{\cZ} ~.
\ee
Then the field strength becomes $\cF = F$, the field strength of a vector multiplet\footnote{Throughout this paper, additional abelian vector multiplets, $\cW$, are described by a similar  two-form 
field strength with the same constraints except with $\cZ$ replaced with $\cW$.}
\begin{subequations}
\begin{align}
F_a{}_\b^j &= \frac{\ri}{2} (\s_a)_\b{}^\gd \bar{\nabla}_\gd^j \bar{\cZ} \ , \qquad 
F_a{}^\bd_j = - \frac{\ri}{2} (\s_a)_\g{}^\bd \nabla^\g_j \cZ \ , \\
F_{ab} &= - \frac{1}{8} (\s_{ab})_{\a\b} ( \nabla^{\a \b} \cZ + 4 W^{\a\b} \bar{\cZ})
+ \frac{1}{8} (\tilde{\s}_{ab})_{\ad\bd}  (\bar{\nabla}^{\ad \bd} \bar{\cZ} + 4 \bar{W}^{\ad\bd} \cZ) \ .
\end{align}
\end{subequations}
This is the standard procedure for gauging the central charge.

Another choice for the superfield is
\be M = \ri \cZ e^{-\ri L} \ ,
\ee
where $L$ is a real primary superfield not annihilated by the central charge and with zero dilatation and $\rm U(1)$ weight. 
In this case the components of $\cF$ were found in \cite{Novak} to be
\begin{align}
\cF_{z}{}_\a^i &= \ri \bm \nabla_\a^i L \ , \quad \cF_z{}^\ad_i = - \ri \bar{\bm \nabla}^\ad_i L \ , \non\\
\cF_a{}^j_\b &= \hf e^{- \ri L} (\s_a)_\b{}^\ad \bar{\bm \nabla}^j_\ad (\bar{\cZ} e^{2 \ri L}) \ , \quad \cF_a{}^\bd_j = \hf e^{\ri L} (\s_a)_\a{}^\bd \bm \nabla_j^\a (\cZ e^{- 2 \ri L}) \ , \non\\
\cF_{az} &= \frac{1}{8} (\s_a)_{\g\gd} [\bm \nabla^{\g k}, \bar{\bm \nabla}^\gd_k] L \ , \non\\
\cF_{ab} &= - \frac{\ri}{8} (\s_{ab})^{\a\b} e^{\ri L} (\bm \nabla_{\a\b} (\cZ e^{-2 \ri L}) - 4 W_{\a\b} \bar{\cZ} ) + \HC \ ,
\end{align}
where $L$ is constrained by
\begin{align}\label{eq_varVT1}
{\bm \nabla}_\a^{(i} \bar {\bm \nabla}_\ad^{j)} L = 0~, \qquad
\re^{\ri L} {\bm \nabla}^{ij} (\cZ \re^{-2 \ri L}) + \re^{-\ri L} \bar {\bm \nabla}^{ij} (\bar\cZ \re^{2 \ri L}) = 0 \ .
\end{align}
Introducing an additional constraint, the simplest version of which is
\be e^{-\ri L} \bm \nabla^{ij} (\cZ e^{2 \ri L}) + e^{\ri L} \bar{\bm \nabla}^{ij} (\bar{\cZ} e^{-2 \ri L}) = 0
\ee
guarantees the existence of a gauge two-form and ensures that $L$ describes a multiplet with $8+8$ degrees of freedom, known as the variant vector-tensor multiplet \cite{Novak}. In the 
rigid supersymmetric limit it reduces to the multiplet constructed by Theis \cite{Theis1, Theis2}.


\section{Superform formulation for vector-tensor multiplets} \label{superform}

As we saw in the previous section, allowing the central charge gauge field to have non-trivial action under the central charge lead us to new possibilities that are not fixed by the 
Bianchi identities. We may fix these extra degrees of freedom by choosing additional constraints. In light of the role that the central charge multiplet plays for the 
VT multiplet, we now turn to fixing these degrees of freedom by demanding the existence of a multiplet with a one-form and a two-form gauge potential.

In this section we will make use of a superform formulation and examine the possibilities allowed by the constraints. The formulation will be a generalization of the one in \cite{Novak},
where superform formulations for both known versions of the VT multiplet in conformal supergravity was given. Our approach has a number of advantages. In particular, it makes 
manifest the existence of two gauge fields (the one-form and the two-form), without the need to go to components and it allows the constraints on superfields to be derived from 
the superform Bianchi identities. Superform formulations are also very useful tools for component reduction (see section \ref{compActions}), where the closure of supersymmetry 
transformations of the component fields are guaranteed by the Bianchi identitites.


\subsection{One-form geometry}

In addition to the central charge gauge one-form we introduce a new gauge one-form, $\tilde{\cV} := E^A \tilde{\cV}_A$, which is also not required to be annihilated by 
the central charge. Its transformation law is defined as (compare with section \ref{GCC})
\be \d \tilde{\cV} = \L \D \tilde{\cV} + \rd \tilde{\G} \ , \quad \D \tilde{\G} = 0 \ ,
\ee
with $\tilde{\G}$ the gauge parameter associated with $\tilde{\cV}$. The corresponding field strength $\tilde{\cF} = \hf E^B E^A \tilde{\cF}_{AB}$ is given by
\be
\tilde{\cF} = \bm \nabla \tilde{\cV} \ , \quad \tilde{\cF}_{AB} = 2 \bm \nabla_{[A} \tilde{\cV}_{B \} } - T_{AB}{}^C \tilde{\cV}_C
\ee
and transforms homogeneously
\be \d \tilde{\cF} = \L \D \tilde{\cF} \ .
\ee
As in \cite{Novak} we may extend the field strength to $\tilde{\cF}_{\hat{A} \hat{B}} = (\tilde{\cF}_{AB}, \tilde{\cF}_{zA} := \D \tilde{\cV}_A)$ and verify that the Bianchi identities may be written 
as
\be \bm \nabla_{[\hat{A}} \tilde{\cF}_{\hat{B}\hat{C}\}} - T_{[\hat{A}\hat{B}}{}^{\hat D} \tilde{\cF}_{|\hat{D}|\hat{C}\}} = 0 \ . \label{FBI}
\ee

We now impose the constraints on the field strength
\begin{align}
\tilde{\cF}_\a^i{}_\b^j = - 2 \eps_{\a\b} \eps^{ij} \bar{N} \ , 
\quad \tilde{\cF}^\ad_i{}^\bd_j = 2 \eps^{\ad \bd} \eps_{ij} N \ , \quad \tilde{\cF}_\a^i{}^\bd_j = 0 \ , \label{Fconstraints}
\end{align}
where $N$ is a primary superfield with dimension 1 and $\rm U(1)$ charge $-2$. The Bianchi identities \eqref{FBI} may be solved giving the field strength components
\begin{align}
\tilde{\cF}_{z}{}_\a^i &= \frac{1}{\bar{M}} \bm \nabla_\a^i \bar{N} \ , \non\\
\tilde{\cF}_a{}_\b^j &= - \frac{\ri}{2} (\s_a)_{\b \ad} ( \bar{\bm \nabla}^{\ad j} \bar{N} - \frac{\bar{M}}{M} \bar{\bm \nabla}^{\ad j} N) \ , \non\\
\tilde{\cF}_{az} &= \frac{\ri}{8} (\s_a)_{\g \gd} \big( \bm \nabla^{\g k} (\frac{1}{M} \bar{\bm \nabla}^\gd_k N) + \bar{\bm \nabla}^\gd_k (\frac{1}{\bar{M}} \bm \nabla^{\g k} \bar{N} )\big) \ , \non\\
\tilde{\cF}_{ab} &= - \frac{1}{8} (\s_{ab})^{\a\b} \Big( \bm \nabla_{\a\b} N - 2 \bm \nabla_\a^k \big(\frac{M}{\bar{M}}\big) \bm \nabla_{\b k} \bar{N} 
- \frac{M}{\bar{M}} \bm \nabla_{\a\b} \bar{N} + 4 \bar{N} W_{\a\b}\Big) + \HC
\end{align}
and the superfield constraints
\begin{subequations} \label{1formconst}
\begin{align}
\bm \nabla_\a^{(i} \Big(\frac{1}{M} \bar{\bm \nabla}_{\bd}^{j)} N\Big) + \bar{\bm \nabla}_\bd^{(i} \Big(\frac{1}{\bar{M}} \bm \nabla_\a^{j)} \bar{N}\Big) &= 0 \ , \label{1formconst1} \\
\bar{N} {\bm \nabla}^{ij} (\frac{M}{\bar{M}}) + {\bm \nabla}^{ij} N - {\bm \nabla}^{ij} (\frac{\bar{N} M}{\bar{M}}) 
&= N \bar{\bm \nabla}^{ij} (\frac{\bar{M}}{M}) + \bar{\bm \nabla}^{ij} \bar{N} - \bar{\bm \nabla}^{ij} (\frac{N \bar{M}}{M}) \ . \label{1formconst2}
\end{align}
\end{subequations}

These results represent new general couplings of the large vector multiplet to a one-form and generalizes the one-form geometry in \cite{Novak} for
both versions of the VT multiplet.


\subsection{Two-form geometry}

The above system of coupled one-forms contain too many component fields. In order to constrain the component fields we demand the existence of a two-form, 
and hence provide a general framework for both versions of the VT multiplet. Similar to 
the analysis in \cite{Novak}, we introduce a gauge two-form, $B = \hf E^B E^A B_{AB}$, with corresponding field strength
\be
H := \bm \nabla B - \hf \eta_{IJ} \tilde{\cV}^I \bm \nabla \tilde{\cV}^J \ ,
\ee
where we couple a number of one-forms, $\tilde{\cV}^I$, to the two-form via the coupling constants $\eta_{IJ}$.\footnote{Here $\eta_{IJ}$ is not required to be symmetric since it was not 
chosen to be symmetric in \cite{Claus3}.} 
The transformation law of $B$ is defined to be
\be \d B = \L \D B + \hf \eta_{IJ} \tilde{\G}^I \rd \tilde{\cV}^J + \rd \Xi \ , \quad \D \X = 0 \ ,
\ee
where the one-form $\Xi$ is the gauge parameter for $B$. As in \cite{Novak} we may extend the field strength to 
$H_{\hat{A} \hat{B} \hat{C}} = (H_{ABC}, H_{zAB} := \D B_{AB} + \eta_{IJ} \tilde{\cV}^I_{[A} \tilde{\cF}^J_{|z| B\}})$, which transforms 
homogeneously
\be \d H_{\hat{A} \hat{B} \hat{C}} = \L \D H_{\hat{A} \hat{B} \hat{C}}
\ee
and satisfies the Bianchi identity
\be
\bm \nabla_{[\hat{A}} H_{\hat{B}\hat{C}\hat{D}\}} - \frac{3}{2} T_{[\hat{A}\hat{B}}{}^\cE H_{|\cE|\hat{C}\hat{D} \}} 
+ \frac{3}{4} \eta_{IJ} \tilde{\cF}^I_{[\hat{A}\hat{B}} \tilde{\cF}^J_{\hat{C}\hat{D}\}} = 0 \ . \label{HBI}
\ee

We now impose constraints on the field strength similar to those in \cite{Novak}
\begin{align} H_\a^i{}_\b^j{}_\g^k &= H^\ad_i{}^\bd_j{}^\gd_k
= H_\a^i{}_\b^j{}^\gd_k = H_\a^i{}^\bd_j{}^\gd_k =  0 \ , \non\\
H_{a}{}_\b^j{}_\g^k &= H_{a}{}^\bd_j{}^\gd_k = 0 \ , \quad H_a{}_\b^j{}^\gd_k = - 2 \ri \d^j_k (\s_a)_\b{}^\gd \tilde{H} \ ,
\end{align}
where $\tilde{H}$ is some real superfield. Without choosing the form of $M$ or $N$ we may analyze \eqref{HBI} to find the components of $H_{\hat{A} \hat{B} \hat{C}}$:
\begin{align}
H_{a \underline{\b} \underline{\g}} =& H_{a \underline{\bd} \underline{\gd}} = 0 \ , \quad H_a{}_\b^j{}^\gd_k = - 2 \ri \d^j_k (\s_a)_\b{}^\gd \tilde{H} \ , \non\\
H_z{}_\a^i{}_\b^j =& - \eps^{ij} \eps_{\a \b} \eta_{IJ} \frac{\bar{N}^I \bar{N}^J}{\bar{M}} \ , \quad H_z{}^\ad_i{}^\bd_j = \eps_{ij} \eps^{\ad \bd} \eta_{IJ} \frac{N^I N^J}{M} \ , \quad H_z{}_\a^i{}^\bd_j = 0 \ , \non\\
H_{a z}{}_\b^j =& - \frac{\ri}{4} \eta_{IJ} (\s_a)_{\b \ad} \Big(\frac{2}{M}\bar{\bm \nabla}^{\ad j} N^{(I} \bar{N}^{J)} - \frac{1}{M \bar{M}} \bar{\bm \nabla}^{\ad j} M \bar{N}^I \bar{N}^J 
- \bar{\bm \nabla}^{\ad j} (\frac{\bar{N}^I \bar{N}^J}{\bar{M}}) \Big) \ , \non\\
H_{ab}{}_\g^k =& 2 (\s_{ab})_\g{}^\a \bm \nabla_\a^k \tilde{H} \ , \quad H_{ab}{}^\gd_k = 2 (\tilde{\s}_{ab})^\gd{}_\ad \bar{\bm \nabla}^\ad_k \tilde{H} \ , \non\\
H_{ab z} =& - \frac{1}{16} \eta_{IJ} (\s_{ab})^{\a\b} \Big( \bm \nabla_{\a\b} (\frac{N^I N^J}{M}) + 4 W_{\a\b} \frac{\bar{N}^I \bar{N}^J}{\bar{M}} - \frac{2}{\bar{M}} \bm \nabla_{\a\b} \bar{N}^{(I} N^{J)} \non\\
& + \frac{1}{M \bar{M}} \bm \nabla_{\a\b} \bar{M} N^I N^J - \frac{4}{\bar{M}} \bm \nabla_\a^k N^{(I} \bm \nabla_{\b k} \bar{N}^{J)} + \frac{2}{\bar{M}} \bm \nabla_\a^k \bar{M} \bm \nabla_{\b k} (\frac{N^I N^J}{M}) \non\\
&+ \frac{2 M}{\bar{M}^2} \bm \nabla_\a^k \bar{N}^I \bm \nabla_{\b k} \bar{N}^J \Big) + \HC \ , \non\\
H_{abc} =& - \frac{1}{32} \eps_{abcd} (\s^d)^{\a\ad} \Big( 2 [\bm \nabla_\a^k, \bar{\bm \nabla}_{\ad k}] \tilde{H} \non\\
&+ \eta_{IJ} M \bar{\bm \nabla}_{\ad}^k (\frac{\bar{M}}{M}) \big( \frac{2}{\bar{M}} \bm \nabla_{\a k} \bar{N}^I N^J - \frac{1}{M \bar{M}} \bm \nabla_{\a k} \bar{M} N^I N^J - \bm \nabla_{\a k} \big(\frac{N^I N^J}{M}\big) \big) \non\\
&+ \eta_{IJ} (\bar{\bm \nabla}_{\ad}^k \bar{N}^I - \frac{\bar{M}}{M} \bar{\bm \nabla}_\ad^k N^I) (\bm \nabla_{\a k} N^J - \frac{M}{\bar{M}} \bm \nabla_{\a k} \bar{N}^J) \Big) + \HC
\end{align}
In addition to the constraints imposed from the one-form geometry we also find the constraints:
\begin{subequations}
\begin{align}
\tilde{H} =& - \frac{1}{2} \eta_{IJ} M \bar{M} Y^I Y^J \ , \quad Y^I := \frac{1}{2 \ri} (\frac{N^I}{M} - \frac{\bar{N}^I}{\bar{M}}) \ , \\
0 &= \eta_{IJ} G^{IJ i j} \ , \non\\
- 4 G^{IJ ij} =& \frac{\ri}{2} \bm \nabla^{ij} (\frac{N^I N^J}{M}) + \frac{\ri}{2 M \bar{M}} \bm \nabla^{ij} \bar{M} N^I N^J - \frac{\ri}{\bar{M}} \bm \nabla^{ij} \bar{N}^{(I} N^{J)} \non\\
& - \frac{2 \ri}{\bar{M}} \bm \nabla^{\g (i} N^{(I} \bm \nabla_{\g}^{j)} \bar{N}^{J)} + \frac{\ri}{\bar{M}} \bm \nabla^{\g (i} \bar{M} \bm \nabla_\g^{j)} (\frac{N^I N^J}{M}) \non\\
&+ \frac{\ri M}{\bar{M}^2} \bm \nabla^{\g (i} \bar{N}^I \bm \nabla^{j)}_\g \bar{N}^J + \HC \ , \label{Hconst1} \\
0 =& \eta_{IJ} [ \bar{\bm \nabla}^{\ad (k} M \bm \nabla^{\a l)} \bar{N}^{(I} + \bm \nabla^{\a (k} \bar{M} \bar{\bm \nabla}^{\ad l)} N^{(I}] Y^{J)} \ . \label{Hconst2}
\end{align}
\end{subequations}

The last constraint \eqref{Hconst2} is quite restrictive when it comes to the choice of $M$ and $N$. It holds if
\be Y^I = \frac{1}{2 \ri} (\frac{N^I}{M} - \frac{\bar{N}^I}{\bar{M}}) = 0
\ee
or
\be \bar{\bm \nabla}^{\ad (k} M \bm \nabla^{\a l)} \bar{N}^I + \bm \nabla^{\a (k} \bar{M} \bar{\bm \nabla}^{\ad l)} N^I = 0 \label{poss2} \ .
\ee
However the case where $Y^I = 0$ is degenerate since $N^I$ drops out of all components of the two-form field strength $H_{ABC}$ and is unconstrained by the two-form geometry. 
The other possibility \eqref{poss2} is satisfied if either $M$ or $N^I$ are chiral or if $N^I/M$ is a real constant. However if $N^I/M$ is a real constant then $Y^I = 0$. Thus we are left 
with two cases to consider, one where $M$ is chiral and the other where $N$ is chiral.

If $M$ is chiral the large vector multiplet reduces to the usual vector multiplet, $M = \cZ$. We may then represent any of the $N^I$ as\footnote{Here we suppress the index $I$ to 
avoid awkward notation.}
\be
N = \cW (R + \ri \tilde{L}) \ ,
\ee
with $R$ and $\tilde{L}$ real superfields and $\cW$ a vector multiplet. It can be seen that the superfield $R$ drops out of the $H_{ABC}$ components and out of the constraints 
\eqref{1formconst1}, \eqref{1formconst2} and \eqref{Hconst1}.\footnote{Although the one-form and its field strength has non-trivial dependence on $R$.} Here we wish to 
describe multiplets with $8+8$ degrees of freedom and therefore we will choose to freeze $R$ to a constant. After freezing $R$, in terms of the superfield constraints, there are two distinct choices for $N$
\be {\rm (i)} \ N = \ri \cW \tilde{L} \ , \qquad {\rm (ii)} \ N = \cW
\ee
up to shifts in $\tilde{L}$ and $\cW$. However, the choice $\rm (i)$  leads to the same form of constraints on $\tilde{L}$ as the choice
\be N = \ri \cZ L \ , \quad L = (\frac{\cW}{\cZ} + \frac{\bar{\cW}}{\bar{\cZ}}) \tilde{L} \ .
\ee
Therefore we may set $N^I = (\ri \cZ L, \cW^{\hat{I}})$ and doing so recovers the constraints for the VT multiplet of \cite{Claus3}
\begin{subequations}  \label{constraintsType1}
\begin{align}
\bm \nabla_\a^{(i} \bar{\bm \nabla}_\bd^{j)} L = 0 \ , \\
\bm \nabla^{ij} (\cZ L) + \bar{\bm \nabla}^{ij}(\bar{\cZ} L) - L \bm \nabla^{ij} \cZ = 0 \ , \\
\eta_{IJ} G^{IJ ij} = 0 \ ,
\end{align}
\end{subequations}
with
\be G^{IJ}{}^{ij} = \hf [\ri \cZ \bm \nabla^{ij} Y^{(I} Y^{J)} + 2 \ri \cZ \bm \nabla^{\g (i} Y^{(I} \bm \nabla_\g^{j)} Y^{J)} + 2 \ri Y^{(I} \bm \nabla^{\g (i} Y^{J)} \bm \nabla_\g^{j)} \cZ] + \HC \ ,
\ee
which appeared in \cite{KN, BN}.\footnote{Here we adopt the convention $\eta_{\hat{I}1} = 0$ in \cite{Claus3}, although it may be more natural to define $\eta_{IJ}$ to be symmetric.}

On the other hand, if $N^I$ are chiral then the relations \eqref{1formconst} constrain them to be reduced chiral. Now we represent $M$ as 
\be M = \ri \cZ R e^{- \ri L} \ ,
\ee
with $R$ and $L$ real superfields. Similar to the previous case, $R$ formally drops out of the $H_{ABC}$ components and out of the constraints 
\eqref{1formconst1}, \eqref{1formconst2} and \eqref{Hconst1}. This time we simply make the choice $R = 1$. Then letting 
$N^I = (\cZ, \cW^{\hat{I}})$ recovers the variant VT multiplet of \cite{Novak}:
\begin{subequations} \label{constraintsType2}
\begin{align}
\bm \nabla_\a^{(i} \bar{\bm \nabla}_\bd^{j)} L &= 0 \ , \\
e^{\ri L}\bm \nabla^{ij} (\cZ e^{- 2 \ri L}) + e^{- \ri L} \bar{\bm \nabla}^{ij} (\bar{\cZ} e^{2 \ri L}) &= 0 \ , \\
e^{- \ri L}\bm \nabla^{ij} (\cZ K e^{2 \ri L}) + e^{\ri L} \bar{\bm \nabla}^{ij} (\bar{\cZ} \bar{K} e^{- 2 \ri L}) &= 0 \ , \quad K \equiv \eta_{11} + \eta_{1\hat{I}} \frac{\cW^{\hat{I}}}{\cZ}
+ \eta_{\hat{I}\hat{J}} \frac{\cW^{\hat{I}} \cW^{\hat{J}}}{\cZ^2} \ .
\end{align}
\end{subequations}

An important consequence of our analysis is that one cannot 
have both kinds of VT multiplets coupled to each other. As a result the two cases from 
here on will be classified as the type I and the type II VT multiplets respectively. Their corresponding superfield Lagrangians are given in \cite{KN, BN, Novak}.


\section{Vector-tensor multiplet in components} \label{compActions}

In the past $\cN = 2$ superconformal tensor calculus \cite{sct_rules, BdeRdeW, sct_structure} has been very popular in constructing component actions 
for matter multiplets in supergravity, {\it e.g.} \cite{sct_lagrangians, deWPV, Claus3}. However as highlighted by the complexity 
of the results in \cite{Claus3}, a superfield formulation is desired due its simpler form. Moreover, it proves advantageous to have superform formulations for the 
supermultiplets. Besides being simpler, superform formulations are useful in identifying covariant field strengths and have an important role to play for 
the ectoplasm (or superform) approach \cite{Ectoplasm, GGKS}. In \cite{BN} the superspace formulations for supergravity of \cite{KLRT-M08, Butter4D} were used to efficiently construct 
component actions entirely from superspace, with superforms for the vector and tensor multiplets facilitating the component reduction.\footnote{The superspace 
formulation of \cite{KLRT-M08} may be derived by a gauge fixing of \cite{Butter4D}.} In this section we 
will make use of the techniques of \cite{BN} and the action principle in \cite{BKN} to derive new component results for VT multiplets.


\subsection{Weyl multiplet}

The Weyl multiplet \cite{sct_rules,  BdeRdeW, sct_structure} may be described by the superspace formulation of \cite{Butter4D}. It contains a set of one-forms: the 
vierbein $e_m{}^a$, the gravitino $\psi_m{}^\a_i$, the $\rm{U}(1)_{R}$ and $\rm{SU}(2)_{R}$ gauge fields $A_m$ and $\phi_m{}^{ij}$ and 
the dilatation gauge field $b_m$. These are simply the component projections of corresponding superforms
\begin{align}
e_m{}^a &:= E_m{}^a| \ , \quad \psi_m{}^\a_i := 2 E_m{}^\a_i| \ , \quad \psib_m{}^i_\ad := 2 E_m{}^i_\ad| \ , \non\\
A_m &:= \Phi_m| \ , \quad \phi_m{}^{i j} := \Phi_m{}^{ij}| \ ,\quad b_m := B_m| ~,
\end{align}
where we define the component projection of a superfield, $V(z)$ by
\be V(z)| := V(z)|_{\theta_i = \bar{\theta}^i = 0} \ .
\ee
The component gauge fields for Lorentz, special conformal and $S$-supersymmetry transformations are denoted by 
$\omega_m{}^{ab}$, $\phi_m{}^i_\a$ and $\frak{f}_m{}^a$ respectively:
\begin{align}
\omega_m{}^{ab} := \Omega_m{}^{ab}| \ , \quad  
\phi_m{}^i_\a := 2 \frak F_m{}^i_\a| \ ,
\quad \bar{\phi}_m{}_i^\ad := 2 \frak F_m{}_i^\ad |~, \quad
\frak{f}_m{}^a := \frak{F}_m{}^a| \ .
\end{align}
The constraints on the superspace curvatures imply that they are composite objects (see appendix \ref{CompIdentities}). Finally, the torsion component $W_{\a\b}$ provides
additional non-gauge component fields
\begin{subequations}
\begin{gather}
W_{ab}^{+} := (\s_{ab})^{\a\b} W_{\a\b}| \ , \quad W_{ab}^{-} := - (\tilde{\s}_{ab})_{\ad\bd} \bar{W}^{\ad\bd}| \ ,\\
\S^{\a i} := \frac{1}{3} \nabla_\b^i W^{\a\b}| \ , \quad \bar{\S}_{\ad i} := - \frac{1}{3} \bar{\nabla}^\bd_i \bar{W}_{\ad\bd} | \ ,\\
D := \frac{1}{12} \nabla^{\a\b} W_{\a\b}| = \frac{1}{12} \bar{\nabla}_{\ad\bd} \bar{W}^{\ad\bd}| \ .
\end{gather}
\end{subequations}
In what follows (as in \cite{BN}) we may drop the component projection operator on $W_{\a\b}$ when it is clear from the context that we are referring to its component projection.


\subsection{Deformed linear multiplet and action principle}

The linear multiplet \cite{Sohnius} is an off-shell representation of $\cN = 2$ supersymmetry with central charge. In supergravity it was used to construct the component action 
for the type I VT multiplet \cite{Claus3}. However, a modification is needed 
when the central charge is gauged in a non-standard way (such as for the type II VT multiplet). It requires a deformed linear multiplet \cite{BKN}, 
which generalizes the usual linear multiplet in the presence of a background large vector multiplet.

The deformed linear multiplet is described by a real symmetric superfield, $\cL^{ji} = \cL^{ij} = (\cL_{ij})^*$ constrained by
\be \widetilde{\bm \nabla}{}_\a^{(i} \cL^{jk)} = 0 \ , \quad \widetilde{\bar{\bm \nabla}}{}^\ad_{(i} \cL_{jk)} = 0 \ ,
\ee
where the tilded derivatives are defined as
\begin{align}
\widetilde{\bar{\bm \nabla}}{}_\a^i = \bar M^{-1} {\bm \nabla}_\a^i \bar M~, \qquad
\widetilde{\bar{\bm \nabla}}{}^\dalpha_i = M^{-1} \bar{\bm \nabla}^\dalpha_i M~.
\end{align}
Its component fields are given by
\begin{subequations}
\begin{alignat}{2}
\ell^{ij} &:= \cL^{ij}|~, \\
\c_{\a i} &:= \frac{1}{3} \widetilde{\bm \nabla}{}_\a^j \cL_{i j}| ~, &\quad 
	\bar{\c}^{\ad i} &:= \frac{1}{3} \widetilde{\bar{\bm \nabla}}{}^\ad_j \cL^{ij}|~, \\
F &:=  \frac{1}{12} \widetilde{\bm \nabla}{}^{ij} \cL_{ij}|~, &\quad
\bar{F} &:=  \frac{1}{12} \widetilde{\bar{\bm \nabla}}{}^{ij} \cL_{ij}| \ ,
\end{alignat}
\end{subequations}
with an additional component (a three-form) associated with the superform $\hat{\S}_{5ABC}$ in \cite{BKN}
\begin{subequations}\label{eq_varH}
\begin{align}
    \hat{\S}_{5}{}_\a^i{}_\b^j{}_\g^k &=  \hat{\S}_{5}{}_\a^i{}_\b^j{}^\gd_k =  \hat{\S}_{5a}{}_\b^i{}_\g^j = 0~, \qquad \hat{\S}_{5a}{}_\b^j{}^\gd_k = 2 (\s_a)_\b{}^\gd \cL^j{}_k \ , \\
    \hat{\S}_{5ab}{}_\alpha^i &= \frac{2\ri}{3} (\s_{ab})_\a{}^\g \widetilde {\bm \nabla}{}_\g^k \cL_k{}^i \ , \\
    \hat{\S}_{5abc} &= \frac{\ri}{24} \eps_{abcd} (\tilde \s^d)^{\ad \a} [\widetilde{\bm \nabla}{}_\a^k, \widetilde{\bar{\bm \nabla}}{}_\ad^l] \cL_{kl} \ .
\end{align}
\end{subequations}

Coupling $\hat{\S}_{5ABC}$ to a four-form $\hat{\S}_{ABCD}$ and using the ectoplasm approach leads to an action principle for the deformed 
linear multiplet \cite{BKN}
\begin{align}\label{eq_LinearAction}
S &= \int \rd^4x\, e\, \Big(F \phi + \chi^\a_i \l_\a^i
	+ \frac{1}{8} \ell^{ij} X_{ij} - v_a \tilde{\S}^a
	\eol & \quad
	- \frac{\ri}{2} \psi_{\a \ad}{}^\a_i ( 2 \bar{\chi}^{\ad i} \bar{\phi} + \ell^{ij} \bar{\l}^{\ad}_j )
	+(\s^{cd})_{\g \d} \psi_c{}^\g_k \psi_d{}_l^\d \ell^{kl} \bar{\phi} + {\rm c.c.}\Big)~,
\end{align}
where
\be \tilde{\S}^a = \frac{1}{6} e_m{}^a \eps^{mnpq} \hat{\S}_{5npq}|
\ee
and
\begin{align}
\phi := M| \ , \quad \l{}_\a^i := \bar M \bm \nabla_\a^i \Big(\frac{M}{\bar M}\Big)| \ , \quad
X^{ij} := \bar M \bm \nabla^{ij} \Big(\frac{M}{\bar M}\Big)| = (X_{ij})^{*} \ , \quad v_a = e_a{}^m \cV_m|
\end{align}
are component fields of the large vector multiplet.\footnote{The Levi-Civita tensor $\eps^{mnpq}$ with world indices is defined as $\eps^{mnpq} := \eps^{abcd} e_a{}^m e_b{}^n e_c{}^p e_d{}^q$.} 
The action has the same form as that for a standard linear multiplet \cite{deWitLM}, however 
the supersymmetry transformations of the component fields have been modified.

On a final note we may construct $\tilde{\S}^a$ from the superform $\hat{\S}_{5ABC}$. Making use of the superspace identity
\begin{align}
\hat{\S}_{5mnp} = E_m{}^A E_n{}^B E_p{}^C \hat{\S}_{5ABC} (-)^{ab+ac+bc}
\end{align}
and taking its superspace projection we find
\begin{align}\label{h_eqn}
\tilde{\S}^a = \frak{E}^a +(\s^{ab})_\a{}^\b \psi_b{}^\a_k \c^k_\b - (\tilde\s^{ab})^\ad{}_\bd \bar\psi_b{}^k_\ad \bar\c^\bd_k
	- \hf \eps^{abcd} (\s_b)_\a{}^\ad \psi_c{}^\a_k \bar\psi_d{}^l_\ad \ell^k{}_l \ ,
\end{align}
with
\be \frak{E}^a := \frac{1}{6} \eps^{abcd} \hat{\S}_{5bcd}| = \frac{\ri}{24} (\tilde \s^a)^{\ad \a} [\widetilde{\bm \nabla}{}_\a^k, \widetilde{\bar{\bm \nabla}}{}_\ad^l] \cL_{kl}\lc
\ee
the final component field of the linear multiplet.


\subsection{Type I vector-tensor multiplet in components}

The type I VT multiplet in section \ref{superform}, possesses the superfield structure 
given by equations \eqref{constraintsType1}. Different choices of $\eta_{\hat{I}\hat{J}}$ lead to the non-linear and linear cases discussed in \cite{Claus3}. However it turns 
out that the nonlinear case with Chern-Simons terms ($\eta_{1 \hat{I}} = 0$) \cite{KN} is equivalent to the most general case by a superfield redefinition \cite{BN}.\footnote{The general case may be derived by making 
the superfield redefinition $L \rightarrow L + c_{I} Y^{I}$, which in terms of the superfield constraints is equivalent to shifts in the field strength $\cF \rightarrow \cF + c_{I} F^{I}$.} 
The constraints for this case may be written as \cite{KN}
\begin{align}
\bm \nabla_\a^{(i} \bar{\bm \nabla}_\ad^{j)} L &= 0 \ , \non\\
\eta_{11} \bm \nabla^{ij} (\cZ L^2) &= 2 \eta_{11} \bar{\cZ} \bar{\bm \nabla}_{\ad}^{(i} L \bar{\bm \nabla}^{\ad j)} L 
+ \frac{1}{4} \eta_{IJ} [ \bm \nabla^{ij}(\frac{\cW^{I} \cW^{J}}{\cZ}) 
- \bar{\bm \nabla}^{ij} (\frac{\bar{\cW}^{I} \bar{\cW}^{J}}{\bar{\cZ}}) ] \ ,
\end{align}
with corresponding Lagrangian
\begin{align} \cL^{ij} =& - \frac{1}{6} \eta_{11} L^3 \bm \nabla^{ij} \cZ + \eta_{11} \cZ L \bm \nabla^i L \bm \nabla^j L + \eta_{11} \bar{\cZ} L \bar{\bm \nabla}^i L \bar{\bm \nabla}^j L 
+ \frac{1}{8} \eta_{IJ} [L \bm \nabla^{ij} (\frac{\cW^I \cW^J}{\cZ}) \non\\
&+ L \bar{\bm \nabla}^{ij} (\frac{\bar{\cW}^I \bar{\cW}^J}{\bar{\cZ}})
- 2 \bm \nabla^{ij} (L \frac{\cW^I \cW^J}{\cZ}) -2 \bar{\bm \nabla}^{ij} (L \frac{\bar{\cW}^I \bar{\cW}^J}{\bar{\cZ}})] \ . \label{T1L}
\end{align}
In principle the above constraints and Lagrangian provides everything we need to perform the component reduction and construct the most general action for the type I VT multiplet. In fact, 
if we are only interested in the bosonic sector then it only requires that we use identities for two spinor derivatives acting on $L$. However, since the bosonic action is given in \cite{Claus3}, here we 
switch off the Chern-Simons terms (by setting $\eta_{\hat{I}\hat{J}} = 0$) and derive the full action for this case. Finally, to match the normalization in \cite{KN} we choose $\eta_{11} = \frac{1}{4}$.

The component fields are defined by
\begin{align}
\ell &:= L \vert \ , \quad \Lambda_\a^i := {\bm\nabla}_\a^i L \vert~,\quad
\bar\Lambda^\ad_i := \bar{\bm\nabla}^\ad_i L \vert~, \quad \ell^{(z)} := \D L\vert \ , \non\\
\tilde{v}_m &:= \tilde{\cV}_m\vert \ , \quad b_{mn} := B_{mn}| \ ,
\end{align}
with the component fields of the large vector multiplet reducing to the component fields of the vector multiplet, $\cZ$:
\begin{align}
\phi = \cZ\vert \ , \quad \l{}_\a^i = \nabla_\a^i \cZ \vert \ , \quad
X^{ij} = \nabla^{ij} \cZ \vert \ , \quad v_m = \cV_m| \ .
\end{align}

The component field strengths are constructed from the projections of their corresponding superspace field strengths
\begin{align}
f_{mn} &:= F_{mn}| = 2 \partial_{[m} v_{n]} \ , \non\\
\tilde{f}_{mn} &:= \tilde{\cF}_{mn}| = 2 (\partial_{[m} + v_{[m} \D) \tilde{\cV}_{n]}\vert = 2 \partial_{[m} \tilde{v}_{n]} + 2 v_{[m} \tilde{f}_{|z|n]} \ , \non\\
\tilde{f}_{zm} &:= \D \tilde{v}_m \equiv \tilde{v}_m^{(z)} \ , \non\\
h_{mnp} &:= 3 [(\partial_{[m} + v_{[m} \D) B_{np]} - \frac{1}{4} \tilde{\cV}_{[m} (\partial_{n} + V_{n} \D ) \tilde{\cV}_{p]}]| \non\\
& \ = 3 [\partial_{[m} b_{np]} - \frac{1}{4} \tilde{v}_{[m} \partial_{n} \tilde{v}_{p]} + v_{[m} h_{|z|np]} ] \ , \non\\
h_{zmn} &:= \D b_{mn} + \frac{1}{4} \tilde{v}_{[m} \tilde{f}_{|z|n]} \ .
\end{align}
Their supercovariant field strengths may be found from superspace. Using the superform components in \cite{Novak} and projecting the identities
\begin{align} F_{mn} &= E_m{}^A E_n{}^B F_{AB}(-)^{ab} \ , \quad \tilde{\cF}_{mn} = E_m{}^A E_n{}^B \tilde{\cF}_{AB}(-)^{ab} \ , \quad \tilde{\cF}_{zm} = E_m^A \tilde{\cF}_{zA} \ , \non\\
H_{mnp} &= E_m{}^A E_n{}^B E_p{}^C H_{ABC} (-)^{ab + ac + bc} \ , \quad H_{zmn} = E_m{}^A E_n{}^B H_{zAB}(-)^{ab} \ ,
\end{align}
yields results corresponding to the those given in \cite{Claus3}.\footnote{Our field strengths are defined slightly differently from \cite{Claus3}, by an 
additional term involving $W_{\a\b}$. The component results for the supercovariant field strengths in \cite{Claus3} 
may also be generated straightforwardly in the general case.} However as in \cite{Claus3}, we will only need to use some of the supercovariant field strengths:
\begin{align}
\hat{F}_{ab} &:= F_{ab}\vert := \hf e_a{}^m e_b{}^n {f}_{mn} - \frac{\ri}{2} (\s_{[a})_\a{}^\ad \psi_{b]}{}^\a_k \bar{\l}{}^k_\ad - \hf \psi_{a}{}^\g_k \psi_{b}{}^k_\g \bar{\phi} + \HC \ , \non\\
\hat{\cF}_{z a} &:= \tilde{\cF}_{za}| = e_a{}^m \tilde{v}_m^{(z)} + \frac{\ri}{2} \psi_a{}^\a_i \L^i_\a - \frac{\ri}{2} \bar{\psi}_a{}^i_\ad \L{}^\ad_i \ , \non\\
\hat{\cF}_{ab} &:= \tilde{\cF}_{ab}| = \hf e_a{}^m e_b{}^n \tilde{f}_{mn} + \hf (\s_{[a}) \psi_{b]}{}^\a_i (2 \bar{\phi} \bar{\L}_\ad^i + \ell \bar{\l}_\ad^i) + \ri \psi_{a}{}^\g_k \psi_b{}^\g_k \bar{\phi} \ell + \HC
\end{align}
The reason why this can be done is because the superform components $H_{zab}$ and $H_{abc}$ can be expressed in terms of the field strength components $\tilde{\cF}_{ab}$ and $\tilde{\cF}_{za}$ 
respectively (see \cite{Novak}).

The action is given in terms of the linear multiplet component fields, derived from \eqref{T1L}. Taking the appropriate covariant derivatives and performing the superspace projection leads to the linear multiplet 
components: {\allowdisplaybreaks
\begin{subequations}
\begin{align}
\ell^{ij} &= - \frac{1}{24} \ell^3 X^{ij} + \frac{1}{4} \phi \ell \L^i \L^j + \frac{1}{4} \bar{\phi} \ell \bar{\L}^i \bar{\L}^j \ , \\
\chi_{\a i} &= - \frac{1}{16} \ell^2 X_{ij} \L_\a^j + \frac{\ri}{12} \ell^3 \nabla_{\a\ad} \bar{\l}^\ad_i
+ \frac{1}{24} \phi \L_\a^j \L_i \L_j \non\\
&+ \frac{1}{8} \bar{\phi} \L_\a^j \bar{\L}_i \bar{\L}_j - \frac{\ri}{2} \ell \hat{\cF}_{\a\b} \L^\b_i - \frac{1}{2} \ell^2 \hat{F}_{\a\b} \L^\b_i \\
&- \frac{1}{2} W_{\a\b} \bar{\phi} \ell^2 \L^\b_i + \frac{1}{8} \ell \l^\b_i \L_\b^j \L_{\a j} - \frac{1}{8} \ell \l_\a^j \L_i \L_j \non\\
&- \frac{1}{4} \bar{\phi} \phi \ell \ell^{(z)} \L_{\a i} + \frac{1}{4} \bar{\phi} \ell (\hat{\cF}_{\a \bd , z} - \ri \bm \nabla_{\a \bd} \ell) \bar{\L}^\bd_i \ , \\
F &= - \frac{1}{4} \bar{\phi} \ell (\hat{\cF}_{az} - \ri \bm \nabla_a \ell) (\hat{\cF}^a{}_z - \ri \bm \nabla^a \ell) - \frac{1}{6} \ell^3 \Box \bar{\phi} + \hf \bar{\phi} \ell^3 D \non\\
&+ \frac{\ell}{2 \phi} (\hat{\cF}_{\a\b} - \ri \ell \hat{F}_{\a\b} - \ri \bar{\phi} \ell W_{\a\b}) (\hat{\cF}^{\a\b} - \ri \ell \hat{F}^{\a\b} - \ri \bar{\phi} \ell W^{\a\b}) - \frac{1}{4} \phi \bar{\phi}^2 \ell (\ell^{(z)})^2 \non\\
&+ \frac{L^3}{128 \phi} X_{ij} X^{ij} + \frac{L^3}{6} \bar{W}^{\ad\bd} (2 \hat{\bar{F}}_{\ad\bd} + \hat{\bar{W}}_{\ad\bd} \bar{\phi}) \non\\
& + \frac{1}{4} \ell^3 \bar{\S}_{\ad k} \bar{\l}^{\ad k} - \frac{\bar{\phi} \ell}{8 \phi} \hat{\cF}_{\a\ad, z} \l^{\a k} \bar{\L}^\ad_k - \frac{\ell}{8} \hat{\cF}_{\a\ad , z} \L^{\a k} \bar{\l}^\ad_k - \frac{\ri \bar{\phi}}{4} \bm \nabla_{\a\ad} \ell \L^{\a k} \bar{\L}^\ad_k \non\\
& + \frac{3 \ri \ell^2}{16} \nabla_{\a\ad} \bar{\l}^{\ad k} \L^\a_k - \frac{\ri \bar{\phi} \ell^2}{8 \phi} \nabla_{\a\ad} \l^{\a k} \bar{\L}^\ad_k + \frac{\ri \ell}{8} \bm \nabla_{\a\ad} \ell \L^{\a k} \bar{\l}^\ad_k - \frac{\ri \bar{\phi}}{8 \phi} \bm \nabla_{\a\ad} \ell \l^{\a k} \bar{\L}^\ad_k \non\\
&- \frac{\ri \bar{\phi} \ell}{4} \bm \nabla_{\a\ad} \L^{\a k} \bar{\L}_{\ad k} + \frac{\ri \bar{\phi} \ell}{4} \L^{\a k} \bm \nabla_{\a\ad} \bar{\L}_{\ad k} + \frac{\bar{\phi} \ell}{4} \ell^{(z)} \bar{\l}^\ad_i \bar{\L}_\ad^i + \frac{\ell^2}{16 \phi} X_{ij} \l^i \L^j \non\\
&- \frac{\ell}{16} X_{ij} \L^i \L^j - \frac{\bar{\phi} \ell}{16 \phi} X_{ij} \bar{\L}^i \bar{\L}^j + \frac{\ri \bar{\phi}}{4 \phi} (\hat{\bar{\cF}}^{\ad\bd} + \ri \ell \hat{\bar{F}}^{\ad\bd}) \bar{\L}_{\ad k} \bar{\L}_\bd^k \non\\
& - \frac{\ri}{4} (\hat{\cF}_{\a\b} - 3 \ri \ell \hat{F}_{\a\b} - 2 \ri \bar{\phi} \ell W_{\a\b}) \L^{\a k} \L^\b_k + \frac{\ri \ell}{2 \phi} (\hat{\cF}_{\a\b} - \ri \ell \hat{F}_{\a\b} - \ri \bar{\phi} \ell W_{\a\b}) \l^{\a k} \L^\b_k \non\\
&- \frac{\ri \bar{\phi} \ell}{4 \phi} \nabla_{\a\ad} \phi \L^{\a k} \bar{\L}^\ad_k + \frac{\ri}{4} \ell \nabla_{\a\ad} \bar{\phi} \L^{\a k} \bar{\L}^\ad_k \non\\
& + \frac{\ell}{8 \phi} \l^i \L^j \l_i \L_j  + \frac{\ell}{8 \phi} \l^k \l^l \L_k \L_l - \frac{\bar{\phi}}{8 \phi} \bar{\l}_i \bar{\L}_j \bar{\L}^i \bar{\L}^j - \frac{1}{12} \l^i \L^j \L_i \L_j - \frac{\bar{\phi}}{8 \phi} \l^i \L^j \bar{\L}_i \bar{\L}_j \non\\
&- \frac{\bar{\phi}}{8 \phi} \bar{\l}_i \bar{\L}_j \bar{\L}^i \bar{\L}^j - \frac{\phi}{32 \ell} \L_i \L_j \L^i \L^j -  \frac{\bar{\phi}^2}{32 \phi \ell} \bar{\L}_i \bar{\L}_j \bar{\L}^i \bar{\L}^j 
+ \frac{\bar{\phi}}{16 \ell} \L_i \L_j \bar{\L}^i \bar{\L}^j \ , \\
\frak{E}_a &= \hf \phi \bar{\phi} \ell \ell^{(z)} \bm \nabla_a \ell - \hf (\ell \hat{\cF}^b{}_{z} + \ri \ell \bm \nabla^b \ell + \hf (\s^b)^\g{}_\gd \L^k_\g \bar{\L}^\gd_k) (\hat{\cF}_{ab}^{+} - \ri \ell \hat{F}_{ab}^{+} - \ri W_{ab}^{+} \bar{\phi} \ell) \non\\
&- \hf (\ell \hat{\cF}^b{}_{z} - \ri \ell \bm \nabla^b \ell + \hf (\s^b)^\g{}_\gd \L^k_\g \bar{\L}^\gd_k) (\hat{\cF}_{ab}^{-} + \ri \ell \hat{F}_{ab}^{-} + \ri W_{ab}^{-} \phi \ell) \non\\
&- \frac{1}{6} \ell^3 \nabla^b (2 \hat{F}_{ab}^{-} + W_{ab}^{-} \phi) - \frac{1}{6} \ell^3 \nabla^b (W_{ab}^{+} \bar{\phi}) - \frac{\ri}{8} (\s_a)_\a{}^\ad \ell^3 \S^{\a k} \bar{\l}_{\ad k} 
- \frac{\ri}{16} (\s^a)^{\a\ad} X_{ij} \L^i_\a \bar{\L}_\ad^j \non\\
&+ \frac{1}{4} \bar{\phi} \ell \bm \nabla_a \L^\gd_k \L_\gd^k - \frac{1}{4} \phi \ell \bm \nabla_a \L_\g^k \L^\g_k 
+ \frac{1}{4} (\s_{ab})^{\a\b} \ell \nabla^b \phi \L_\a^k \L_{\b k} - \frac{1}{4} (\tilde{\s}_{ab})^{\ad\bd} \ell \nabla^b \bar{\phi} \bar{\L}_{\ad k} \bar{\L}_{\bd}^k \non\\
&- \frac{1}{8} \ell^2 \nabla_a \l_\g^k \L^\g_k + \frac{1}{8} \ell^2 \nabla_a \bar{\l}^\gd_k \bar{\L}_\gd^k
+ \frac{1}{4} (\s_{ab})^{\a\b} \ell^2 \nabla^b \l_\a^k \L_{\b k} - \frac{1}{4} (\tilde{\s}_{ab})^{\ad\bd} \ell^2 \nabla^b \bar{\l}_{\ad k} \bar{\L}_\bd^k \non\\
& + \frac{\ri \ell}{8} (\hat{\cF}_{az} + \ri \bm \nabla_a \ell) \l_\g^k \L^\g_k + \frac{\ri \ell}{8} (\hat{\cF}_{az} - \ri \bm \nabla_a \ell) \bar{\l}^\gd_k \bar{\L}^k_\gd \displaybreak \non\\
& - \frac{\ri \ell}{4} (\s_{ab})^{\a\b} (\hat{\cF}^b{}_z + \ri \bm \nabla^b \ell) \l_\a^k \L_{\b k} - \frac{\ri \ell}{4} (\tilde{\s}_{ab})^{\ad\bd} (\hat{\cF}^b{}_z - \ri \bm \nabla^b \ell) \bar{\l}^\ad_k \bar{\L}^{\bd k} \non\\
& - \frac{\ri}{16} (\s_a)^{\a\ad} \l_i \L_j \L^j_\a \bar{\L}^i_\ad - \frac{\ri}{16} (\s^a)^{\a\ad} \bar{\l}^i \bar{\L}^j \L_{\a i} \bar{\L}_{\ad j} \non\\
& - \frac{\ri}{16} (\s^a)^{\a\ad} \l_\a^i \bar{\L}^j_\ad \L_i \L_j + \frac{\ri}{16} \bar{\l}_\ad^i \L_\a^j \bar{\L}_i \bar{\L}_j \ ,
\end{align}
\end{subequations}}
where for compactness we have defined a number of composite objects. We define for terms arising from vector derivatives of the vector multiplet fields:
\begin{subequations}
\begin{align}
\nabla_a \phi &\equiv \nabla'_a \phi - \hf (\psi_a{}_j \lambda^{j})~, \\
\nabla_b \bar\lambda{}^{\dalpha}_j &\equiv
     \nabla'_b \bar\lambda{}^\dalpha_j
     + 2 \bar \phi \bphi_b{}^\dalpha_j 
     - \ri \psi_{b \beta j} \nabla^{\dalpha \beta} \bar \phi
     + \frac{1}{4} \bar{\psi}_b{}^{\dalpha k} X_{kj}
     + 2 \bar{\psi}_{b \dbeta j} \hat{F}^{\dbeta \dalpha}
     + \bar{\psi}_{b \dbeta j} \bar W^{\dbeta \dalpha} \phi~, \\
\Box \bar\phi &\equiv
     \nabla'^a \nabla_a \bar\phi
	+ 2 \frak{f}_a{}^a \bar \phi
	- \frac{\ri}{2} (\phi_m{}^j \sigma^m \bar\lambda_j)
     \eol &\quad
     + \frac{\ri}{4} (\psi_{m j} \sigma^m)_\dalpha \bar W^{\dalpha \dbeta} \bar\lambda_{\dbeta}^j
     - \frac{3\ri}{4} (\psi_{m j} \sigma^m \bar\S^j) \bar\phi
     - \hf \bar{\psi}_{a \dbeta}^j \nabla^a \bar\lambda^\dbeta_j \ , \\
\nabla_a \hat{F}_{\a\b} &= \nabla_a' \hat{F}_{\a\b} - \phi_a{}_{(\a}^k \l_{\b) k} + \frac{1}{4} \psi_a{}_{(\a k} (\ri \nabla_{\b)}{}^\gd \bar{\l}^k_\gd + 2 \S_{\b)}^k \bar{\phi}) \non\\
&- \frac{1}{4} \psi_a{}^\g_k (2 \Psi_{(\a\b}{}_{\g)}^k + \ri \Psi_{(\a \ad}{}^{\ad k} W_{\b \g)}) \bar{\phi} - \bar{\psi}_a{}^k_\gd (\ri \nabla_{(\a}{}^\gd \l_{\b) k} - W_{\a\b} \bar{\l}^\gd_k) \ ,
\end{align}
\end{subequations}
as well as for the vector-tensor fields
\begin{subequations}
\begin{align} 
\bm \nabla_a \ell &\equiv \bm \nabla'_a \ell - \hf (\psi_a{}_j \L^{j}) -  \hf (\bar{\psi}_a{}^j \bar{\L}_{j}) \ , \\
\bm \nabla_a \L_\a^i &\equiv \bm \nabla_a' \L_\a^i - \hf \psib_a{}_\bd^i \big( \hat{\cF}_\a{}^\bd{}_{, z} + \ri \bm \nabla_\a{}^\bd \ell \big) \non\\
&+ \frac{1}{4} \psi_{a \a j} \big( - \frac{1}{\ell} \L^i \L^j + \frac{\bar{\phi}}{\phi \ell} \bar{\L}^i \bar{\L}^j - \frac{2}{\phi} \l^{(i} \L^{j)} - \frac{\ell}{2 \phi} X^{ij} \big) \non\\
&- \hf \psi_a{}^{\b i} \big( \frac{2 \ri}{\phi} \hat{\cF}_{\a\b} + \frac{2 \ell}{\phi} \hat{F}_{\a\b} + \frac{2 \bar{\phi} \ell}{\phi} W_{\a\b} - \frac{1}{\phi} \l_{(\a} \L_{\b)} \big) + \hf \psi_a{}^i_\a \bar{\phi} \ell^{(z)} \ .
\end{align}
\end{subequations}
We also define
\be \nabla_a W_{\a\b}| = \nabla_a' W_{\a\b} \ .
\ee
The derivative $\bm \nabla_a' = \nabla_a' + v_a \D $, with $\nabla_a'$ given in appendix \ref{CompIdentities}, is covariant with respect to Lorentz, dilatation, $\rm U(1)_R$ and $\rm SU(2)_R$ transformations. 
Finally, from the constraints one can derive the following useful identity {\allowdisplaybreaks
\begin{align}
\phi \D \L_\a^i &= \ri \bm \nabla_{\a\ad} \bar{\L}^{\ad i} - \l_\a^i \ell^{(z)} - \frac{1}{4 \ell^2} \L_{\a j} \bar{\L}^i \bar{\L}^j - \frac{1}{2 \ell} \hat{\cF}_{\a\ad , z} \bar{\L}^{\ad i} \non\\
&+ \frac{\ri}{2 \ell} \bm \nabla_{\a\ad} \ell \bar{\L}^{\ad i} + \frac{\phi}{4 \bar{\phi} \ell^2} \L_{\a j} \L^i \L^j - \frac{\ri}{\bar{\phi} \ell} \hat{\cF}_{\a\b} \L^{\b i} - \frac{1}{\bar{\phi}} \hat{F}_{\a\b} \L^{\b i} \non\\
& - W_{\a\b} \L^{\b i} - \frac{\phi}{2 \ell} \ell^{(z)} \L_\a^i + \frac{\ri}{\bar{\phi}} \nabla_{\a\bd} \bar{\phi} \bar{\L}^{\bd i} - \frac{1}{2 \bar{\phi}} \hat{\cF}_{\a\bd , z} \bar{\l}^{\bd i} \non\\
&+ \frac{\ri}{2 \bar{\phi}} \bm \nabla_{\a\bd} \ell \bar{\l}^{\bd i} + \frac{1}{8 \bar{\phi}} X^{ij} \L_{\a j} + \frac{\ri \ell}{2 \bar{\phi}} \nabla_{\a\ad} \bar{\l}^{\ad i} \non\\
&+ \frac{1}{4 \bar{\phi} \ell} \l^{\b i} \L_\a^j \L_{\b j} + \frac{1}{4 \bar{\phi} \ell} \l_{\a j} \L^i \L^j \ ,
\end{align}}
which may be used to make explicit the terms containing $v_a$.

The bosonic sector of the above results can be shown to agree with \cite{Claus3}.\footnote{There is however a conventional difference, 
which changes where the auxiliary field $D$ emerges (see \cite{Butter4D}).}


\subsection{Type II vector-tensor multiplet in components}

The type II VT multiplet possesses the superfield structure given by equations \eqref{constraintsType2}. The corresponding Lagrangian 
was constructed in \cite{Novak}
\begin{align}
\cL^{ij} &= \frac{\ri}{2} e^{-\ri L} \bm \nabla^{ij} (\cZ K L e^{2 \ri L}) - e^{- \ri L} \bar{\cZ} \bar{K} \bar{\bm \nabla}^i L \bar{\bm \nabla}^j L - \frac{1}{4} e^{- \ri L} \bar{\bm \nabla}^{ij}(\bar{\cZ} \bar{K}) + \HC
\end{align}
To simplify the component reduction we will look at the pure case where $\eta_{11} = 1$, $\eta_{1\hat{I}} = \eta_{\hat{I} \hat{J}}= 0$, which represents a supergravity extension 
of the VT multiplet of Theis \cite{Theis2}. In this case the constraints may be written as
\begin{align}
\bm \nabla_\a^{(i} \bar{\bm \nabla}_\ad^{j)}  L =& 0 \ , \\
\bm \nabla^{ij} L =& - \frac{2}{\tan 2L} \bm \nabla^i L \bm \nabla^j L - \frac{2 \bar{\cZ}}{\cZ \sin 2 L} \bar{\bm \nabla}^i L \bar{\bm \nabla}^j L - \frac{2}{\cZ} \bm \nabla^{(i} \cZ \bm \nabla^{j)} L \non\\
&+ \frac{1}{2 \cZ \tan L} \bm \nabla^{ij} \cZ \ ,
\end{align}
with corresponding Lagrangian
\begin{align} 
\cL^{ij} =& - 2 e^{\ri L} \cZ \bm \nabla^i L \bm \nabla^j L - 2 e^{-\ri L} \bar{\cZ} \bar{\bm \nabla}^i L \bar{\bm \nabla}^j L - \cos L \bm \nabla^{ij} \cZ + \ri L e^{- \ri L} \bm \nabla^{ij} (\cZ e^{2 \ri L}) \non\\
=& \ 2 \cZ \Big(\frac{2 L e^{-\ri L}}{\sin 2L} - e^{\ri L}\Big) \bm \nabla^i L \bm \nabla^j L + 2 \bar{\cZ} \Big(\frac{2 L e^{\ri L}}{\sin 2L} - e^{- \ri L}\Big) \bar{\bm \nabla}^i L \bar{\bm \nabla}^j L \non\\
&- \Big( \cos L + \frac{L}{\sin L} \Big) \bm \nabla^{ij} \cZ \ .
\end{align}

The component fields are
\begin{align}
\ell &:= L \vert \ , \quad \Lambda_\a^i := {\bm\nabla}_\a^i L \vert~,\quad
\bar\Lambda^\ad_i := \bar{\bm\nabla}^\ad_i L \vert~, \quad \ell^{(z)} := \D L\vert \ , \non\\
v_m &:= \cV_m\vert \ , \quad b_{mn} := B_{mn}| \ ,
\end{align}
with the component fields of the vector multiplet, $\cZ$, defined by\footnote{The tildes signify the differing roles $\cZ$ play for the two types of VT multiplets.}
\begin{align}
\tilde{\phi} = \cZ\vert \ , \quad \tilde{\l}{}_\a^i = \nabla_\a^i \cZ \vert \ , \quad
\tilde{X}^{ij} = \nabla^{ij} \cZ \vert \ , \quad \tilde{v}_m = \tilde{V}_m| \ .
\end{align}

The component field strengths and their corresponding supercovariant field strengths are constructed as in the previous subsection. The component field strengths are
\begin{align}
\tilde{f}_{mn} &:= F_{mn}| = 2 \partial_{[m} \tilde{v}_{n]} \ , \non\\
f_{mn} &:= \cF_{mn}| = 2 (\partial_{[m} + v_{[m} \D) \cV_{n]}\vert = 2 \partial_{[m} \tilde{v}_{n]} + 2 v_{[m} f_{zn]} \ , \non\\
f_{zm} &:= \D v_m \equiv v_m^{(z)} \ , \non\\
h_{mnp} &:= 3 [(\partial_{[m} + v_{[m} \D) B_{np]} - \frac{1}{4} \tilde{\cV}_{[m} (\partial_{n} + V_{n} \D) \tilde{\cV}_{p]}]| \non\\
& \ = 3 [\partial_{[m} b_{np]} - \frac{1}{4} \tilde{v}_{[m} \partial_{n} \tilde{v}_{p]} + v_{[m} h_{|z|np]} ] \ , \non\\
h_{zmn} &:= \D b_{mn} \ ,
\end{align}
with corresponding supercovariant field strengths
\begin{align}
\hat{F}_{ab} &:= F_{ab}\vert := \hf e_a{}^m e_b{}^n {\tilde{f}}_{mn} - \frac{\ri}{2} (\s_{[a})_\a{}^\ad \psi_{b]}{}^\a_k \tilde{\bar{\l}}{}^k_\ad - \hf \psi_{a}{}^\g_k \psi_{b}{}^k_\g \tilde{\bar{\phi}} + \HC \ , \non\\
\hat{\cF}_{za} &:= \cF_{za}| = e_a{}^m v_m^{(z)} - \frac{\ri}{2} \psi_a{}^\a_i \L^i_\a + \frac{\ri}{2} \bar{\psi}_a{}^i_\ad \bar{\L}^\ad_i \ , \non\\
\hat{\cF}_{ab} &:= \cF_{ab}| = \hf e_a{}^m e_b{}^n f_{mn} + \hf (\s_{[a})_\a{}^\ad \psi_{b]}{}^\a_i (\tilde{\bar{\l}}_\ad^i + 2 \ri \bar{\L}_\ad^i) e^{\ri \ell} + \ri \psi_{a}{}^\g_k \psi_b{}_\g^k \tilde{\bar{\phi}} e^{\ri \ell} + \HC \ , \non\\
\hat{H}_{zab} &:= \hat{H}_{zab}| = \hf e_a{}^m e_b{}^n h_{zmn} - \frac{1}{4} (\s_{[a})_\a{}^\ad \psi_{b]}{}^\a_i (\tilde{\bar{\l}}_\ad^i - 2 \ri \bar{\L}_\ad^i) e^{- \ri \ell} - \frac{\ri}{2} \tilde{\psi}_{a}{}^\g_k \psi_b{}_\g^k \tilde{\bar{\phi}} e^{-\ri \ell} + \HC \ , \non\\
\hat{H}_{abc} &:= H_{abc}| = \hf e_a{}^m e_b{}^n e_c{}^p h_{mnp} + \frac{3}{4} (\s_{[ab})_\a{}^\g \psi_{b]}{}^\a_i \tilde{\bar{\phi}} \Big((\cos 2\ell + 1) \tilde{\l}_\g^i - 2 \tilde{\phi} \sin (2 \ell) \L_\g^i\Big) \non\\
&\qquad \qquad \qquad +\frac{3}{4} (\s_{[a})_\a{}^\bd \psi_{b}{}^\a_k \bar{\psi}_{c]}{}_\bd^k \tilde{\phi} \tilde{\bar{\phi}} (\cos 2 \ell + 1) + \HC
\end{align}
Similar to the previous case we will only need the expressions for $\hat{F}_{ab}$, $\hat{\cF}_{za}$ and $\hat{\cF}_{ab}$. 

The type II VT multiplet possesses more complicated component structure, which is highlighted by the appearance of trigonometric functions. Therefore for simplicity sake we 
provide results for the bosonic sector of the action. The component fields of the linear multiplet are: {\allowdisplaybreaks
\begin{subequations}
\begin{align} 
\ell^{ij} =& \ 2 \tilde{\phi} \Big(\frac{2 \ell e^{-\ri \ell}}{\sin 2\ell} - e^{\ri \ell}\Big) \L^i \L^j + 2 \tilde{\bar{\phi}} \Big(\frac{2 \ell e^{\ri \ell}}{\sin 2\ell} - e^{- \ri \ell}\Big) \bar{\L}^i \bar{\L}^j \non\\
&- \Big( \cos \ell + \frac{\ell}{\sin \ell} \Big) X^{ij} \ , \\
\c_{\a i} &= \Big(- \frac{2 e^{2 \ri \ell} + 1}{4 \sin \ell} + \frac{\ell e^{-\ri \ell}}{2 \sin^2 \ell} \Big) X_{ij} \L_\a^j \non\\
&- 2 \Big( \frac{2 \ell e^{- \ri \ell}}{\sin 2\ell} - e^{\ri \ell} \Big) \Big(2 \ri \hat{F}_{\a\b} + \ri W_{\a\b} \tilde{\bar{\phi}} (1+ e^{2 \ri \ell})- 2 e^{\ri \ell} \hat{\cF}_{\a\b} \Big) \L^\b_i \non\\
& + 2 \ri \tilde{\phi} \tilde{\bar{\phi}} e^{\ri \ell} \Big( \frac{2 \ell e^{- \ri \ell}}{\sin 2 \ell} - e^{\ri \ell} \Big) \ell^{(z)} \L_{\a i}\non\\
&- 2 \tilde{\bar{\phi}} \Big(\frac{2 \ell e^{\ri \ell}}{\sin 2\ell} - e^{- \ri \ell} \Big) (\hat{\cF}_{\a\ad, z} + \ri \bm \nabla_{\a\ad}' \ell) \bar{\L}^\ad_i \non\\
& + 2 \ri (\cos \ell + \frac{\ell}{\sin \ell}) \bm \nabla_{\a\ad} \bar{\bm \nabla}^{\ad}_i \tilde{\bar{\phi}} + \rm{3 \ fermion \ terms} \ , \\
F &= \frac{1}{32 \tilde{\phi} \sin \ell \tan \ell} \Big( \frac{2\ell e^{- \ri \ell}}{\sin \ell} - 2 e^{2 \ri \ell} + 1 \Big) \tilde{X}^{ij} \tilde{X}_{ij} \non\\
&- \frac{1}{\tilde{\phi}} \Big( \frac{2 \ell e^{- \ri \ell}}{\sin 2\ell} - e^{\ri \ell} \Big) (2 \ri \hat{F}_{\a\b} + \ri W_{\a\b} \tilde{\bar{\phi}} (1+ e^{2 \ri \ell}) - 2 e^{\ri \ell} \hat{\cF}_{\a\b}) \non\\
&\qquad \qquad \times (2 \ri \hat{F}^{\a\b} + \ri W^{\a\b} \tilde{\bar{\phi}} (1+ e^{2 \ri \ell}) - 2 e^{\ri \ell} \hat{\cF}^{\a\b}) \non\\
&+2 \tilde{\phi} \tilde{\bar{\phi}}^2 e^{2 \ri \ell} \Big(\frac{2 \ell e^{-\ri \ell}}{\sin 2\ell} - e^{\ri \ell}\Big) (\ell^{(z)})^2 \non\\
&- 2 \tilde{\bar{\phi}} \Big( \frac{2 \ell e^{\ri \ell}}{\sin 2 \ell} - e^{- \ri \ell} \Big) (\hat{\cF}_{a z} + \ri \bm \nabla_a' \ell) (\hat{\cF}^a{}_{z} + \ri \bm \nabla'^a \ell) \non\\
& - 4 (\cos \ell + \frac{\ell}{\sin \ell}) \Box \tilde{\bar{\phi}} + 4 (\cos \ell + \frac{\ell}{\sin \ell}) \bar{W}_{\ad\bd} (2 \hat{\bar{F}}^{\ad\bd} + \bar{W}^{\ad\bd} \tilde{\phi}) \non\\
& + 12 \tilde{\bar{\phi}} (\cos \ell + \frac{\ell}{\sin \ell}) D + \rm{fermion \ terms} \ , \\
\frak{E}_a &= - 2 \ri \Big( \frac{2 \ell e^{- \ri \ell}}{\sin 2\ell} - e^{\ri \ell} \Big) (2 \ri \hat{F}_{ab}^+ + \ri W_{ab}^{+} \tilde{\bar{\phi}} (1 + e^{2 \ri \ell}) - 2 e^{\ri \ell} \hat{\cF}_{ab}^{+}) (\hat{\cF}^b{}_z - \ri \bm \nabla'^b \ell) \non\\
& - 2 \ri \Big( \frac{2 \ell e^{\ri \ell}}{\sin 2\ell} - e^{- \ri \ell} \Big) (2 \ri \hat{F}_{ab}^- + \ri W_{ab}^{-} \tilde{\phi} (1 + e^{- 2 \ri \ell}) - 2 e^{- \ri \ell} \hat{\cF}_{ab}^{-}) (\hat{\cF}^b{}_z + \ri \bm \nabla'^b \ell) \non\\
& + 4 \tilde{\phi} \tilde{\bar{\phi}} \Big( \frac{2 \ell}{\sin 2 \ell} - \cos 2 \ell \Big) \ell^{(z)} \hat{\cF}_{a z} - 4 \tilde{\phi} \tilde{\bar{\phi}} \sin (2 \ell) \ell^{(z)} \bm \nabla'_a \ell \non\\
& - 4 (\cos \ell + \frac{\ell}{\sin \ell}) \bm \nabla^b (2 \hat{F}_{ab}^- + W_{ab}^- \tilde{\phi})- 4 (\cos \ell + \frac{\ell}{\sin \ell}) \bm \nabla^b (W_{ab}^+ \tilde{\bar{\phi}}) \non\\
&+ \rm{fermion \ terms} \ .
\end{align}
\end{subequations}}

The composite objects are defined as in the previous section, but with the component fields of $\cZ$ defined with a tilde. Finally the component fields of the large vector multiplet appearing in the 
action principle \eqref{eq_LinearAction} are
\begin{align}
\phi &= \ri \tilde{\phi} e^{-\ri \ell} \ , \quad \l_\a^i = \ri e^{- \ri \ell}  \tilde{\l}_\a^i + 2 e^{- \ri \ell} \tilde{\phi} \L_\a^i \ , \non\\
X^{ij} &= \frac{1}{\sin \ell} \tilde{X}^{ij} - 4 \frac{e^{\ri \ell}}{\sin 2 \ell} \tilde{\phi} \L^i \L^j - 4 \frac{e^{- \ri \ell}}{\sin 2 \ell} \tilde{\bar{\phi}} \bar{\L}^i \bar{\L}^j \ .
\end{align}


\section{Conclusion} \label{discus}

One of the main goals of this paper was to exhaust possible versions of the VT multiplet in conformal superspace. This was motivated by the numerous papers exploring possible deformations 
of the original VT multiplet. Generalizing the superform formulations of \cite{Novak} to a suitable framework for all known VT multiplets, we found two distinct cases. Interestingly, these 
two cases correspond to generalizations of the possible deformations in flat superspace. Moreover, the 
possibility of making use of the variant VT multiplet to construct an alternative off-shell formulation for the VT multiplet of \cite{Claus3} was ruled out.

The second main goal was to construct component actions for VT multiplets using superspace techniques. We found the construction of the component 
action from a superfield Lagrangian to be efficient, with superform formulations facilitating the component reduction. Most notably the full component action for the standard non-linear VT 
multiplet was derived together with the bosonic sector for the supergravity extension of the VT multiplet of Theis \cite{Theis2}. These represent completely new results and it would be 
interesting if the type II VT multiplet turns out to have applications in string theory. If this turns out to be the case then our superform formulation would be indispensable.


\noindent
{\bf Acknowledgements:}\\
I would like to thank Sergei Kuzenko for supervision and helpful comments on the manuscript. 
This work is supported by an Australian postgraduate award.


\appendix


\section{Conformal superspace} \label{conformalSpace}

This appendix contains a brief summary of conformal superspace of \cite{Butter4D}.\footnote{We use the conventions of \cite{BN}, which follow 
closely the conventions of \cite{Ideas}.} Consider a curved 4D $\cN = 2$ superspace $\cM^{4|8}$ parametrized by local bosonic $(x)$ 
and fermionic $(\q, \bar{\q})$ coordinates 
$z^M = (x^m, \ \q^\mu_\imath, \ \bar{\q}_{\dot{\mu}}^\imath)$, where
$m = 0, 1, \cdots, 3,$ $\mu = 1, 2$, $\dot{\mu} = 1, 2$ and $\imath = \1, \2$. The
Grassmann variables $\q^\mu_\imath$ and $\bar{\q}_{\dot{\mu}}^\imath$ are related
to each other by complex conjugation: $\overline{\q^\mu_\imath} = \bar{\q}^{\dot{\mu} \imath}$.
The covariant derivatives
$\nabla_A = (\nabla_a, \nabla_\a^i , \bar{\nabla}^\ad_i)$ have the form
\begin{align} \nabla_A &= E_A + \hf \Omega_A{}^{ab} M_{ab} + \Phi_A{}^{ij} J_{ij} + \ri \Phi_A Y 
+ B_A \mathbb{D} + \frak{F}_{A}{}^B K_B \non\\
&= E_A + \Omega_A{}^{\b\g} M_{\b\g} + \bar{\Omega}_A{}^{\bd\gd} \bar{M}_{\bd\gd} + \Phi_A{}^{ij} J_{ij} + \ri \Phi_A Y 
+ B_A \mathbb{D} + \frak{F}_{A}{}^B K_B \ .
\end{align}
Here $E_A = E_A{}^M(z) \partial_M$ is the supervielbein, with $\partial_M = \partial/\partial z^M$,
$J_{kl} = J_{lk}$ are generators of the group $\rm SU(2)_R$,
$M_{ab}$ are the Lorentz generators, $Y$ is the generator of the chiral rotation
group $\rm U(1)_R$, and $K^A = (K^a, S^\a_i, \bar{S}_\ad^i)$ are the special
superconformal generators. The one-forms $\Omega_A{}^{bc}$, $\Phi_A{}^{kl}$, $\Phi_A$, $B_A$ and
$\frak{F}_A{}^B$ are the corresponding connections.

The generators act on the covariant derivatives as
\begin{align}
[M_{ab}, \nabla_c ] &= 2 \eta_{c [a} \nabla_{b]}~, \quad [M_{ab}, \nabla_\a^i] = (\s_{ab})_\a{}^\b \nabla_\b^i ~, \quad
[M_{ab}, \bar\nabla^\ad_i] = (\tilde{\s}_{ab})^\ad{}_\bd \bar\nabla^\bd_i~, \non\\
[J_{ij}, \nabla_\a^k] &= - \d^k_{(i} \nabla_{\a j)} ~,\quad
[J_{ij}, \bar\nabla^\ad_k] = - \ve_{k (i} \bar\nabla^{\ad}_{j)}~, \non \\
[Y, \nabla_\a^i] &= \nabla_\a^i ~,\quad [Y, \bar\nabla^\ad_i] = - \bar\nabla^\ad_i~,  \non \\
[\mathbb{D}, \nabla_a] &= \nabla_a ~, \quad
[\mathbb{D}, \nabla_\a^i] = \hf \nabla_\a^i ~, \quad
[\mathbb{D}, \bar\nabla^\ad_i] = \hf \bar\nabla^\ad_i ~ .
\end{align}

Finally, the algebra of $K^A$ with $\nabla_B$ is given by
\begin{align}
[K^a, \nabla_b] &= 2 \delta^a_b \mathbb{D} + 2 M^{a}{}_b ~,\non \\
\{ S^\a_i , \nabla_\b^j \} &= 2 \d^j_i \d^\a_\b \mathbb{D} - 4 \d^j_i M^\a{}_\b 
- \d^j_i \d^\a_\b Y + 4 \d^\a_\b J_i{}^j ~,\non \\
\{ \bar{S}^i_\ad , \bar{\nabla}^\bd_j \} &= 2 \d^i_j \d^\bd_\ad \mathbb{D} 
+ 4 \d^i_j \bar{M}_\ad{}^\bd + \d^i_j \d_\ad^\bd Y - 4 \d_\ad^\bd J^i{}_j ~,\non \\
[K^a, \nabla_\b^j] &= -\ri (\s^a)_\b{}^\bd \bar{S}_\bd^j \ , \quad [K^a, \bar{\nabla}^\bd_j] = 
-\ri ({\s}^a)^\bd{}_\b S^\b_j ~, \non \\
[S^\a_i , \nabla_b] &= \ri (\s_b)^\a{}_\bd \bar{\nabla}^\bd_i \ , \quad [\bar{S}^i_\ad , \nabla_b] = 
\ri ({\s}_b)_\ad{}^\b \nabla_\b^i \ ,
\end{align}
where all other (anti-)commutations vanish.

The covariant derivatives obey the (anti-)commutation relations: {\allowdisplaybreaks
\begin{subequations}\label{CSGAlgebra}
\begin{align}
\{ \nabla_\a^i , \nabla_\b^j \} &= 2 \ve^{ij} \ve_{\a\b} \bar{W}_{\gd\dd} \bar{M}^{\gd\dd} + \hf \ve^{ij} \ve_{\a\b} \bar{\nabla}_{\gd k} \bar{W}^{\gd\dd} \bar{S}^k_\dd 
- \hf \ve^{ij} \ve_{\a\b} \nabla_{\g\dd} \bar{W}^\dd{}_\gd K^{\g \gd}~, \\
\{ \bar{\nabla}^\ad_i , \bar{\nabla}^\bd_j \} &= - 2 \ve_{ij} \ve^{\ad\bd} W^{\g\d} M_{\g\d} + \frac{1}{2} \ve_{ij} \ve^{\ad\bd} \nabla^{\g k} W_{\g\d} S^\d_k 
- \frac{1}{2} \ve_{ij} \ve^{\ad\bd} \nabla^{\g\gd} W_{\g}{}^\d K_{\d \gd}~, \\
\{ \nabla_\a^i , \bar{\nabla}^\bd_j \} &= - 2 \ri \d_j^i \nabla_\a{}^\bd~, \\
[\nabla_{\a\ad} , \nabla_\b^i ] &= - \ri \ve_{\a\b} \bar{W}_{\ad\bd} \bar{\nabla}^{\bd i} - \frac{\ri}{2} \ve_{\a\b} \bar{\nabla}^{\bd i} \bar{W}_{\ad\bd} \mathbb{D} 
- \frac{\ri}{4} \ve_{\a\b} \bar{\nabla}^{\bd i} \bar{W}_{\ad\bd} Y + \ri \ve_{\a\b} \bar{\nabla}^\bd_j \bar{W}_{\ad\bd} J^{ij}
	\eol & \quad
	- \ri \ve_{\a\b} \bar{\nabla}_\bd^i \bar{W}_{\gd\ad} \bar{M}^{\bd \gd} - \frac{\ri}{4} \ve_{\a\b} \bar{\nabla}_\ad^i \bar{\nabla}^\bd_k \bar{W}_{\bd\gd} \bar{S}^{\gd k} 
	+ \frac{1}{2} \ve_{\a\b} \nabla^{\g \bd} \bar{W}_{\ad\bd} S^i_\g
	\eol & \quad
	+ \frac{\ri}{4} \ve_{\a\b} \bar{\nabla}_\ad^i \nabla^\g{}_\gd \bar{W}^{\gd \bd} K_{\g \bd}~, \\
[ \nabla_{\a\ad} , \bar{\nabla}^\bd_i ] &=  \ri \d^\bd_\ad W_{\a\b} \nabla^{\b}_i + \frac{\ri}{2} \d^\bd_\ad \nabla^{\b}_i W_{\a\b} \mathbb{D} 
- \frac{\ri}{4} \d^\bd_\ad \nabla^{\b}_i W_{\a\b} Y + \ri \d^\bd_\ad \nabla^{\b j} W_{\a\b} J_{ij}
	\eol & \quad
	+ \ri \d^\bd_\ad \nabla^{\b}_i W^\g{}_\a M_{\b\g} + \frac{\ri}{4} \d^\bd_\ad \nabla_{\a i} \nabla^{\b j} W_\b{}^\g S_{\g j} - \hf \d^\bd_\ad \nabla^\b{}_\gd W_{\a\b} \bar{S}^{\gd}_i
	\eol & \quad
	+ \frac{\ri}{4} \d^\bd_\ad \nabla_{\a i} \nabla^\g{}_\gd W_{\b\g} K^{\b\gd} ~.
\end{align}
\end{subequations}}
The complex superfield $W_{\a\b} = W_{\b\a}$ and its complex conjugate
${\bar{W}}_{\ad \bd} := \overline{W_{\a\b}}$ are superconformally primary,
$K_A W_{\a\b} = 0$, and obey the additional constraints
\begin{align}
\bar{\nabla}^\ad_i W_{\b\g} = 0~,\qquad
\nabla_{\a\b} W^{\a\b} &= \bar{\nabla}^{\ad\bd} \bar{W}_{\ad\bd} ~,
\end{align}
where
\begin{align}
\nabla_{\a\b} := \nabla_{(\a}^k \nabla_{\b) k} \ , \quad \bar{\nabla}^{\ad\bd} := \nabla^{(\ad}_k \nabla^{\bd) k} \ .
\end{align}

When solving Bianchi identities the following list of non-vanishing torsion components are useful:
\begin{align}
T_\a^i{}^\bd_j{}^a &= - 2 \ri \d^i_j (\s^a)_\a{}^\bd \ , \non\\
T_a{}^j_\b{}^k_\gd &= - \frac{\ri}{2} \eps^{jk} (\s_a)_\b{}^\bd \bar{W}_{\bd\gd} \ , \quad T_a{}^\bd_j{}^\g_k = - \frac{\ri}{2} \eps_{jk} (\s_a)_\b{}^\bd W^{\b\g} \ , \non\\
T_{ab}{}^\g_k &= \frac{1}{4} (\s_{ab})^{\a\b} \nabla^\g_k W_{\a\b} \ ,  \quad T_{ab}{}_\gd^k = \frac{1}{4} (\tilde{\s}_{ab})^{\ad\bd} \bar{\nabla}_\gd^k \bar{W}_{\ad\bd} \ .
\end{align}


\section{Component results from superspace} \label{CompIdentities}

In this appendix, we summarize component results of the conformal superspace in \cite{Butter4D}. The notation here differs from \cite{Butter4D} 
and is summarized in \cite{BN} (see also \cite{Ideas}).

The supercovariant derivative is derived from the component projection of $\nabla_m$
\begin{align}
e_m{}^a \nabla_a \lc &= \nabla'_m
	- \hf \psi_m{}^\g_k \nabla_\g^k\lc - \hf \bar{\psi}_m{}^\gd_k \bar{\nabla}_\gd^k\lc
	+ \frak{f}_m{}^b K_b + \hf \phi_m{}^k_\g S^\g_k + \hf \phi_m{}_k^\gd \bar{S}_\gd^k~,
\end{align}
where
\be \nabla_m'  := \left(\partial_m + \frac{1}{2} \omega_m{}^{ab} M_{ab}
		+ \phi_m{}^{ij} J_{ij}
		+ \ri A_m Y
		+ b_m \mathbb D\right) \ .
\ee
So that the projection of the covariant derivatives with Lorentz indices may be written as
\begin{align}
\nabla_a\lc = \nabla_a'
	- \hf \psi_a{}^\g_k \nabla_\g^k\lc - \hf \bar{\psi}_a{}^\gd_k \bar{\nabla}_\gd^k\lc
	+ \frak{f}_a{}^b K_b + \hf \phi_a{}^k_\g S^\g_k + \hf \phi_a{}_k^\gd \bar{S}_\gd^k~,
\end{align}
where
\begin{align}
\nabla_a' := e_a{}^m \nabla_m' \ , \quad \bar{\psi}_a{}^\gd_k := e_a{}^m \bar{\psi}_m{}^\gd_k \ , \quad \frak{f}_a{}^b := e_a{}^m \frak{f}_m{}^a \ , \quad \phi_a{}^i_\a := e_a{}^m \phi_m{}^i_\a \ .
\end{align}

Constraints on the form of the superspace torsion and curvatures imply that the connections $\omega_m{}^{ab}$, $\frak{f}_m{}^a$ and $\phi_m{}^i_\a$ are composite fields built 
out of the other components of the Weyl multiplet (see the torsion analysis of \cite{Butter4D}). The spin connection is given by
\be \omega_{abc} = \hf (\cC_{bca} + \cC_{acb} - \cC_{abc}) - \hf (\cT_{bca} + \cT_{acb} - \cT_{abc}) - b_m e_a{}^m \eta_{ca} - b_m e_c{}^m \eta_{ba} \ ,
\ee
where
\be \cC_{ab}{}^c := 2 e_{[a}{}^n \partial_n e_{b]}{}^m e_m{}^c \ , \quad \cT_{ab}{}^c := \ri (\s^c)_{\g}{}^\dd \psi_{[a}{}^\g_k \bar{\psi}_{b]}{}^k_{\dd} \ .
\ee

The $S$-supersymmetry connection is
\begin{align}
\phi_{\beta \dbeta\,}{}_\alpha^j &=
     \frac{\ri}{12} \bar\Psi_\beta{}^\gd{}_{\alpha \gd\,}{}_\dbeta^j
     + \frac{\ri}{6} \bar\Psi_\alpha{}^\gd{}_{\beta \gd\,}{}_\dbeta^j
     + \frac{\ri}{12} \eps_{\beta \alpha} \bar\Psi_\dbeta{}^\g{}_{\g \gd}{}^{\gd j}
     \eol & \quad
     - \frac{1}{6} \bar W_\dbeta{}^\gd \psi_{\beta \gd\,}{}_\alpha^j
     - \frac{1}{3} \bar W_\dbeta{}^\gd \psi_{\alpha \gd}{}_\beta^j
	+ \frac{\ri}{2} \eps_{\beta \alpha} \bar\S_\bd^j~,
	\\
\bphi_{\beta \dbeta \, \dalpha j} &=
     -\frac{\ri}{12}  \Psi_\dbeta{}^\g{}_{\g \dalpha \,\beta j}
     - \frac{\ri}{6} \Psi_\dalpha{}^\g{}_{\g \dbeta \,\beta j}
     - \frac{\ri}{12} \eps_{\dbeta \dalpha} \Psi_\beta{}^\gd{}_{\g \gd\,}{}^\g_j
     \eol & \quad
     + \frac{1}{6} W_\beta{}^\g \bpsi_{\g \dbeta \dalpha j}
     + \frac{1}{3} W_\beta{}^\g \bpsi_{\g \dalpha \dbeta j}
	+ \frac{\ri}{2} \eps_{\dbeta \dalpha} \S_{\beta j}~,
\end{align}
where we define the gravitino field strength
\be \Psi_{ab}{}^\g_k := e_a{}^m e_b{}^n \Psi_{mn}{}^\g_k \ , \quad \Psi_{mn}{}^\g_k := 2 \nabla_{[m}' \psi_{n]}{}^\g_k  \ .
\ee

The trace of the special conformal connection $\frak{f}_m{}^a$ is (only the trace is required for our calculations):
\begin{align}
\frak{f}_a{}^a &=  -D -\frac{1}{12} \poin \cR
     + \frac{1}{24} \eps^{mnpq} (\bar{\psi}_m{}^j \ts_n \nabla'_p \psi_{q j})
     - \frac{1}{24} \eps^{mnpq} (\psi_{m j} \sigma_n \nabla'_p \bar{\psi}_q{}^j)
     \eol & \quad
     + \frac{\ri}{8} (\psi_{a j} \sigma^a \bar \S^j)
     - \frac{\ri}{8} (\bar{\psi}_a{}^j \ts^a \S_j)
     + \frac{1}{12} W^{ab +} (\bar{\psi}_a{}^j \bar{\psi}_{bj})
     - \frac{1}{12} W^{ab-} (\psi_{a j} \psi_b{}^j)~.
\end{align}
Here $\poin\cR = \poin\cR(e, \omega)$ corresponds to the Poincar\'e version of the Lorentz
curvature
\be \hat{\cR}_{ab}{}^{cd} := e_a{}^m e_b{}^n (\partial_{[m} \omega_{n]}{}^{cd} + \omega_{[m}{}^{cf} \omega_{n] f}{}^d ) \ .
\ee

On a final note the following identity will prove useful:
\be \nabla_{(\a}^k W_{\b\g)} = - 2 \Psi_{(\a\b}{}_{\g)}^k - \ri \Psi_{(\a \ad}{}^{\ad k} W_{\b \g)} \ .
\ee


\begin{footnotesize}

\end{footnotesize}


\begin{thebibliography}{66}


\bibitem{Fayet}
P.~Fayet, ``Fermi-Bose hypersymmetry,''
Nucl.\ Phys.\ B {\bf 113}, 135 (1976).

\bibitem{Sohnius}
  M.~F.~Sohnius,
``Supersymmetry and central charges,''
Nucl.\ Phys.\  B {\bf 138}, 109 (1978).

\bibitem{SSW}
  M.~Sohnius, K.~S.~Stelle and P.~C.~West,
  ``Off-mass-shell formulation of extended supersymmetric gauge theories,''
  Phys.\ Lett.\  B {\bf 92}, 123 (1980);
``Dimensional reduction by Legendre transformation  generates off-shell supersymmetric Yang-Mills theories,''
 Nucl.\ Phys.\  B {\bf 173}, 127 (1980).

\bibitem{SSW2}
  M.~F.~Sohnius, K.~S.~Stelle and P.~C.~West,
 ``Representations of extended supersymmetry,''
in {\it Superspace and Supergravity}, S. W. Hawking and M. Ro\v{c}ek (Eds.), 
Cambridge University Press, Cambridge, 1981, p. 283.

\bibitem{deWKLL}
  B.~de Wit, V.~Kaplunovsky, J.~Louis and D.~L\"ust,
``Perturbative couplings of vector multiplets in N=2 heterotic string  vacua,''
  Nucl.\ Phys.\  B {\bf 451}, 53 (1995)
  [arXiv:hep-th/9504006].

\bibitem{Milewski}
 B.~Milewski,
  ``Superfield  formulation of the N=2 super-Yang-Mills model with central charge,''
Phys.\ Lett.\  B {\bf 112} (1982) 148;
``N=1 superspace formulation of N=2 and N=4 super-Yang-Mills models with central charge,''
Nucl.\ Phys.\  B {\bf 217}, 172 (1983).
  
\bibitem{Claus1}
  P.~Claus, B.~de Wit, M.~Faux, B.~Kleijn, R.~Siebelink and P.~Termonia,
  ``The vector-tensor supermultiplet with gauged central charge,''
  Phys.\ Lett.\  B {\bf 373}, 81 (1996)
  [arXiv:hep-th/9512143].

\bibitem{Claus2}
  P.~Claus, P.~Termonia, B.~de Wit and M.~Faux,
  ``Chern-Simons couplings and inequivalent vector-tensor multiplets,''
  Nucl.\ Phys.\  B {\bf 491}, 201 (1997)
  [arXiv:hep-th/9612203].
  
\bibitem{Claus3}
  P.~Claus, B.~de Wit, M.~Faux, B.~Kleijn, R.~Siebelink and P.~Termonia,
  ``N=2 supergravity Lagrangians with vector-tensor multiplets,''
  Nucl.\ Phys.\  B {\bf 512}, 148 (1998)
  [arXiv:hep-th/9710212].
  
\bibitem{HOW}
  A.~Hindawi, B.~A.~Ovrut and D.~Waldram,
  ``Vector-tensor multiplet in N=2 superspace with central charge,''
  Phys.\ Lett.\  B {\bf 392}, 85 (1997)
  [arXiv:hep-th/9609016].

\bibitem{GHH}
 R.~Grimm, M.~Hasler and C.~Herrmann,
  ``The N=2 vector-tensor multiplet, central charge superspace, and
  Chern-Simons couplings,''
  Int.\ J.\ Mod.\ Phys.\  A {\bf 13}, 1805 (1998)
  [arXiv:hep-th/9706108].

\bibitem{BHO}
  I.~Buchbinder, A.~Hindawi and B.~A.~Ovrut,
  ``A two-form formulation of the vector-tensor multiplet in central charge
  superspace,''
  Phys.\ Lett.\  B {\bf 413}, 79 (1997)
  [arXiv:hep-th/9706216].
  
\bibitem{DKT}
  N.~Dragon, S.~M.~Kuzenko and U.~Theis,
  ``The vector-tensor multiplet in harmonic superspace,''
  Eur.\ Phys.\ J.\  C {\bf 4}, 717 (1998)
  [arXiv:hep-th/9706169].
  
\bibitem{DK}
  N.~Dragon and S.~M.~Kuzenko,
  ``Self-interacting vector-tensor multiplet,''
  Phys.\ Lett.\  B {\bf 420}, 64 (1998)
  [arXiv:hep-th/9709088].

\bibitem{IS}
  E.~Ivanov and E.~Sokatchev,
  ``On nonlinear superfield versions of the vector tensor multiplet,''
  Phys.\ Lett.\  B {\bf 429}, 35 (1998)
  [arXiv:hep-th/9711038].

\bibitem{DIKST}
  N.~Dragon, E.~Ivanov, S.~Kuzenko, E.~Sokatchev and U.~Theis,
  ``N=2 rigid supersymmetry with gauged central charge,''
  Nucl.\ Phys.\  B {\bf 538}, 411 (1999)
  [arXiv:hep-th/9805152].
  
\bibitem{KN}
  S.~M.~Kuzenko and J.~Novak,
  ``Vector-tensor supermultiplets in AdS and supergravity,''
    JHEP {\bf 0112}, 106 (2012) [arXiv:1110.0971 [hep-th]].
  
\bibitem{BN} 
  D.~Butter and J.~Novak,
  ``Component reduction in N=2 supergravity: the vector, tensor, and vector-tensor multiplets,''
  arXiv:1201.5431 [hep-th].
  
\bibitem{Novak}
  J.~Novak,
  ``Superform formulation for vector-tensor multiplets in conformal supergravity,''
  JHEP {\bf 1209}, 060 (2012)
  [arXiv:1205.6881 [hep-th]].
  
\bibitem{ADS} 
  L.~Andrianopoli, R.~D'Auria and L.~Sommovigo,
  ``D=4, N=2 supergravity in the presence of vector-tensor multiplets and the role of higher p-forms in the framework of free differential algebras,''
  Adv.\ Stud.\ Theor.\ Phys.\  {\bf 1}, 561 (2008)
  [arXiv:0710.3107 [hep-th]].

\bibitem{ADST} 
  L.~Andrianopoli, R.~D'Auria, L.~Sommovigo and M.~Trigiante,
  ``D=4, N=2 Gauged Supergravity coupled to Vector-Tensor Multiplets,''
  Nucl.\ Phys.\ B\ {\bf 851}, 1  (2011)
  [arXiv:1103.4813 [hep-th]].
  
\bibitem{KLRT-M08}
	S.~M.~Kuzenko, U.~Lindstr\"om, M.~Ro\v{c}ek and G.~Tartaglino-Mazzucchelli,
	``4D N = 2 supergravity and projective superspace,''
	JHEP {\bf 0809}, 051 (2008)
	arXiv:0805.4683 [hep-th]; 
 ``On conformal supergravity and projective superspace,''
  JHEP {\bf 0908}, 023 (2009)
  [arXiv:0905.0063 [hep-th]].

\bibitem{Butter4D}
  D.~Butter,
  ``N=2 Conformal Superspace in Four Dimensions,''
  JHEP {\bf 1110}, 030 (2011)
  [arXiv:1103.5914 [hep-th]].
  
\bibitem{Theis1}
  U.~Theis,
  ``New N=2 supersymmetric vector tensor interaction,''
  Phys.\ Lett.\  B {\bf 486}, 443 (2000)
  [arXiv:hep-th/0005044].

\bibitem{Theis2}
U.~Theis,  ``Nonlinear vector tensor multiplets revisited,''
Nucl.\ Phys.\  B {\bf 602}, 367 (2001)
  [arXiv:hep-th/0012096].
  
\bibitem{BKN} 
  D.~Butter, S.~M.~Kuzenko and J.~Novak,
  ``The linear multiplet and ectoplasm,''
  JHEP {\bf 1209}, 131 (2012).
  [arXiv:1205.6981 [hep-th]].
  
\bibitem{BS}
P.~Breitenlohner and M.~F.~Sohnius,
``Superfields, auxiliary fields, and tensor calculus for N=2 extended
supergravity,''
Nucl.\ Phys.\  B {\bf 165}, 483 (1980).

\bibitem{deWvHVP3}
  B.~de Wit, J.~W.~van Holten and A.~Van Proeyen,
  ``Central charges and conformal supergravity,''
  Phys.\ Lett.\ B {\bf 95}, 51 (1980).
  
\bibitem{KT}
  S.~M.~Kuzenko and S.~Theisen,
  ``Correlation functions of conserved currents in N=2 superconformal theory,''
  Class.\ Quant.\ Grav.\  {\bf 17}, 665 (2000)
  [hep-th/9907107].
  
\bibitem{HarmSpace} 
  A.~S.~Galperin, E.~A.~Ivanov, V.~I.~Ogievetsky and E.~S.~Sokatchev,
  ``Harmonic superspace,''
  Cambridge, UK: Univ. Pr. (2001) 306 p.
  
\bibitem{Howe}
  P.~S.~Howe,
  ``Supergravity in superspace,''
  Nucl.\ Phys.\  B {\bf 199}, 309 (1982).  
  
\bibitem{Ectoplasm} 
S.~J.~Gates, Jr, ``Ectoplasm has no topology: The prelude,''
in {\it Supersymmetries and Quantum Symmetries},
 J. Wess and E. A. Ivanov (Eds.), Springer, Berlin, 1999, p. 46, arXiv:hep-th/9709104;
``Ectoplasm has no topology,''
 Nucl.\ Phys.\  B {\bf 541}, 615 (1999)
 [arXiv:hep-th/9809056].
 
\bibitem{GGKS}
S.~J.~Gates, Jr., M.~T.~Grisaru, M.~E.~Knutt-Wehlau and W.~Siegel,
``Component actions from curved superspace: Normal coordinates and
ectoplasm,'' Phys.\ Lett.\  B {\bf 421}, 203 (1998)
[hep-th/9711151].

\bibitem{Vor} T. Voronov, ``Geometric integration theory on supermanifolds,''
 Sov. Sci. Rev. C: Math. Phys. {\bf 9},  1 (1992).
 
\bibitem{Hasler} 
  M.~F.~Hasler,
  ``The three-form multiplet in N=2 superspace,''
  Eur.\ Phys.\ J.\ C {\bf 1}, 729 (1998)
  [hep-th/9606076].
  
  
\bibitem{AGHH} 
  G.~Akemann, R.~Grimm, M.~Hasler and C.~Herrmann,
  ``N=2 central charge superspace and a minimal supergravity multiplet,''
  Class.\ Quant.\ Grav.\  {\bf 16}, 1617 (1999)
  [hep-th/9812026].
  
\bibitem{GSW}
  R.~Grimm, M.~Sohnius, J.~Wess,
  ``Extended Supersymmetry and Gauge Theories,''
  Nucl.\ Phys.\  {\bf B133}, 275 (1978).
  


\bibitem{sct_rules}
  B.~de Wit, J.~W.~van Holten, A.~Van Proeyen,
  ``Transformation Rules of N=2 Supergravity Multiplets,''
  Nucl.\ Phys.\  {\bf B167}, 186 (1980).
  
\bibitem{BdeRdeW}
  E.~Bergshoeff, M.~de Roo and B.~de Wit,
  ``Extended conformal supergravity,''
  Nucl.\ Phys.\  B {\bf 182}, 173 (1981).
  
\bibitem{sct_structure}
  B.~de Wit, J.~W.~van Holten, A.~Van Proeyen,
  ``Structure of N=2 Supergravity,''
  Nucl.\ Phys.\  {\bf B184}, 77 (1981).

\bibitem{sct_lagrangians}
  B.~de Wit, P.~G.~Lauwers and A.~Van Proeyen,
  ``Lagrangians Of N=2 Supergravity - Matter Systems,''
  Nucl.\ Phys.\  B {\bf 255}, 569 (1985).
  
\bibitem{deWPV}
B.~de Wit, R.~Philippe and A.~Van Proeyen,
``The improved tensor multiplet in N = 2 supergravity,''
Nucl.\ Phys.\ B {\bf 219}, 143 (1983).

\bibitem{deWitLM} 
  B.~de Wit, J.~W.~van Holten and A.~Van Proeyen,
  ``Central Charges And Conformal Supergravity,''
  Phys.\ Lett.\ B {\bf 95}, 51 (1980).
  

  
\bibitem{Ideas} I.~L.~Buchbinder and S.~M.~Kuzenko,
{\it Ideas and Methods of Supersymmetry and
Supergravity or a Walk Through Superspace},
IOP, Bristol, 1998.

\end{thebibliography}
\end{document}